\documentclass[%
twocolumn,superscriptaddress,
 amsmath,amssymb,
 aps,
prc,
]{revtex4-2}
\usepackage{mathrsfs,amsmath,amssymb,amsthm}
\usepackage{amssymb,fmtcount}
\usepackage{graphicx,epsfig,latexsym,overpic,amssymb,color}
\usepackage{threeparttable}
\usepackage{amsmath} 
\usepackage{float}

\everymath{\displaystyle}
\usepackage{graphicx}
\usepackage{dcolumn}
\begin{document}


\title{ Properties of chiral nucleon-nucleon interaction at N$^3$LO
with high cutoffs studied by local projection
}


\author{H.Y. Shang}
\affiliation{
State Key Laboratory of Nuclear Physics and Technology, School of Physics,
Peking University, Beijing 100871, China
}
\author{R.Z. Hu}
\affiliation{
State Key Laboratory of Nuclear Physics and Technology, School of Physics,
Peking University, Beijing 100871, China
}
\author{J.C. Pei}\email{peij@pku.edu.cn}
\affiliation{
State Key Laboratory of Nuclear Physics and Technology, School of Physics,
Peking University, Beijing 100871, China
}
\affiliation{
Southern Center for Nuclear-Science Theory (SCNT), Institute of Modern Physics, Chinese Academy of Sciences, Huizhou 516000,  China
}
\author{F.R. Xu}\email{frxu@pku.edu.cn}
\affiliation{
State Key Laboratory of Nuclear Physics and Technology, School of Physics,
Peking University, Beijing 100871, China
}
\affiliation{
Southern Center for Nuclear-Science Theory (SCNT), Institute of Modern Physics, Chinese Academy of Sciences, Huizhou 516000,  China
}


\date{\today}

\begin{abstract}
The chiral nucleon-nucleon ($NN$) interaction at high cutoffs has been plagued by the presence of spurious bound states.
In this work, the chiral $NN$ interaction at N$^3$LO is studied by the local projection method as the cutoff increases. 
The evolution of short-range behaviors of pion-exchange interactions and contact interactions is intuitively demonstrated. 
The $P$-channel potentials toward high cutoffs appear to be erratic at short ranges to compromise with phase shifts, while such erratic
behaviors can be avoided in $S$ and $D$ channels. Furthermore, a chiral $NN$ interaction at N$^3$LO is studied at a cutoff of 700 MeV.
 The properties of deuteron and triton are testified with this interaction. 
Such a hard interaction is expected to provide an alternative choice for studies of short-range correlations and high density nuclear matter.
\end{abstract}


\maketitle

\section{Introduction}

The chiral effective field theory ($\chi$EFT) presents a major breakthrough to derive
modern nuclear forces \cite{Chiraleffectivefield,Moderntheorynuclear,Moderntheorynuclear1}. $\chi$EFT provides a perturbation solution of low energy QCD
and maintains its crucial feature of chiral symmetry. Within $\chi$EFT, the nuclear force
invokes the pion degrees of freedoms and many-body forces can be generated in a systematic expansion. 
The three-nucleon forces appear at N$^2$LO and are important for descriptions of nuclear properteis \cite{Threenucleonforcesimplementation}. 
The applications of $\chi$EFT nuclear forces have been very successful in {\it ab initio} calculations
of nuclear structures \cite{tichai2020many,initionocore,Coupledclustercomputationsatomic,inmediumsimilarityrenormalizationa,soma2020self,Continuumthreenucleonforce,initioGamowinmedium,initionocoreGamow,DescriptionProtonDecayingResonance}, 
few-nucleon reactions \cite{EffectThreeNucleonInteractions,Threenucleonforceschiral,initiomanybodycalculations,Lowenergyneutrondeuteronreactionsa,DescriptionProtonDecayingResonance} 
and nuclear matter \cite{ChiralEffectiveFielda,ChiralInteractionsNexttoNexttoNexttoLeadinga,Coupledclustercalculationsnucleonica,GreenFunctionTechniques}. 

The nuclear force based on $\chi$EFT has been conventionally written in terms of ($Q/\Lambda_{\chi}$)$^v$, where $Q$ denotes the momentum
and $\Lambda_{\chi}$ is the chiral symmetry broken scale, and $v$ is the order of the Weinberg power counting \cite{WEINBERG1990,WEINBERG19913}. 
In addition, an explicit cutoff $\Lambda$ in chiral interaction is realized by multiplying a regulator function.
Currently the high quality chiral nuclear interaction up to N$^4$LO has been
developed \cite{Highqualitytwonucleonpotentials,Semilocalmomentumspaceregularized}. 
The  chiral nuclear interactions usually adopt a finite cutoff around 500 MeV and the $NN$ scattering data are fitted  below 300 MeV \cite{ 
Accuratenuclearradii,OptimizedChiralNucleonNucleon,Accuratechargedependentnucleonnucleon,Highqualitytwonucleonpotentials}.  
In the framework of Weinberg power counting, the chiral  interactions are known to have regulator dependence \cite{Improvedchiralnucleonnucleona,Semilocalmomentumspaceregularized}.

In Weinberg's power counting, the
potential is obtained by summing irreducible diagrams
in orders of $Q/\Lambda_{\chi}$, followed by a non-perturbative resummation in the Lippmann-Schwinger equation to infinite
order to yield the amplitude \cite{Renormalizationonepionexchange}. This can be a problem for renormalization of chiral interactions \cite{ProblemRenormalizationChiral,Renormalizationchiraltwopiona}. 
With a high cutoff,  the limit-cycle behaviors of parameters \cite{Singularpotentialslimit} and unphysical deep bound states can emerge in some partial waves,
which are related to the presence of attractive singular potentials. 
Such behaviors have been studied at LO and N$^2$LO before \cite{Renormalizationonepionexchange,Improvingconvergencechiral1}. 
However, the emergence of spurious states at high cutoffs at N$^3$LO have not been studied comprehensively yet. 

The chiral interaction is naturally developed in momentum space. The local projection method
provides a direct visualization of the range-dependent features of the non-local momentum interactions.
This has been successfully applied in studying the softening behavior of nuclear interactions in similarity renormalization group evolution \cite{Localprojectionslowmomentum}. 
This method is expected to be useful to study the cutoff dependence of chiral interactions,
since it is too complicated to study the limit-cycle behaviors and deep bound states at N$^3$LO analytically.
For nuclear interactions, not only the phase shift but also the 
 behaviors of nuclear interactions in coordinate space should be considered. 

It is always curious to explore realistic nuclear forces at high cutoffs although soft interactions are favorable in many-body calculations. 
Actually  $NN$ scattering data in the elastic channel is available up to 3 GeV \cite{Updatedanalysiselastic}. 
The $\Delta$ degree of freedom would play a role at high cutoffs but the role of full chiral interactions with $\Delta$
is not conclusive in actual calculations \cite{LightNucleiSpectraChiral,Nucleonnucleonpotentialsfull,Chiraleffectivefield}.
The role of three-nucleon forces is different in hard $NN$ interactions \cite{Highqualitytwonucleonpotentials}. 
The high energy electron scattering measurement indicates that 
hard nuclear interactions such as Argonne-V18~\cite{Accuratenucleonnucleonpotential} can better describe short-range correlations~\cite{ProgressUnderstandingShortRangea,Probingcorestrong}.
In addition, it would be more realistic to describe nuclear matter at higher densities using  nuclear interactions with high cutoffs. 

In this work,  we study the behaviors of chiral nuclear interactions at N$^3$LO using the local
projection method while the cutoff varies. Thus the emergence of spurious deep bound states can be intuitively
demonstrated, which is difficult to be studied analytically at N$^3$LO. 
The optimization of chiral $NN$ interactions at N$^3$LO is also performed with a high cutoff of 700 MeV. 
The short range features of this interaction are also studied with local projection.
The properties of deuteron and triton are calculated with this hard interaction. 

\section{Methods}
\subsection{The chiral NN potential}
Generally the nuclear potential is the sum of contact terms $V_{\rm ct}$ and 
pion-exchange terms $V_{\rm pe}$ \cite{Chiraleffectivefield}:
\begin{equation}
    V\left({\vec{p}}~', \vec p\right)=V_{\rm ct}\left({\vec p}~', \vec{p}\right)
    +V_{\rm pe}\left({\vec p}~', \vec{p}\right),
\end{equation}
where ${\vec p}~'$ and $\vec{p}$ denote the final and initial nucleon momenta in
the center-of-mass (CMS) system.

In Weinberg's power counting, 
the potential is expanded in terms of powers of small external 
momenta over a large scale as $(Q/\Lambda_\chi)^\nu$. The pion-exchange potential is the sum of various pion-exchange 
contributions.
The 
leading order is abbreviated as LO, and next-to leading order as NLO, etc.
The contact potential is the sum of even polynomials of the momentum as:
\begin{equation}
    \begin{aligned}
    V_{\rm ct}= & C_S+C_1 q^2+C_2 k^2+D_1 q^4+D_2 k^4\\
    &+D_3 q^2 k^2+D_4 \left(\vec{q} \times \vec{k}\right)^2+\cdots,  
    \end{aligned}
\end{equation}
where the momentum transfer is denoted by $\vec{q} \equiv {\vec p}~'-\vec{p}$ and 
the average momentum by $\vec{k} \equiv \left({\vec p}~'+\vec{p}\right)/2$.
In coordinate spaces, the contact potential is associated with delta function and derivatives of delta function.

The specific form of the pion-exchange
component is given in Ref.~\cite
{Peripheralnucleonnucleonscattering,Highqualitytwonucleonpotentials}. For parameters of the axial-vector coupling constant 
$g_A$, pion-decay constant $f_\pi$, pion masses, nucleon masses and 
$\pi N$ low-energy constants (LECs), we adopt the values in Table I and II of 
Ref.~\cite{Highqualitytwonucleonpotentials}. To study the contributions of
different types of diagrams, the two-pion exchange (2PE) terms within $\mathrm{N^3LO}$ ($V_{2 \pi}^{(4)}$)
are further decomposed \cite{Peripheralnucleonnucleonscattering} to facilitate discussions. To be specific, the $2\pi$ contribution is decomposed into 
one-loop and two-loop (tl) contributions. The one-loop contribution includes terms of
$c_i^2$ (football diagram or fd), corrections related to $c_i/M_N$ (cm), and leading relativistic corrections (rc), as given in Appendix A. 
The contact terms under the partial wave representation are 
summerized in Appendix B. The specific charge dependency 
will be discussed  in subsequent sections. The chiral $NN$ interaction can be expressed as
\begin{equation}
    \begin{aligned}
        V_{\mathrm{LO}} & =V_{1 \pi}+V_{\mathrm{ct}}^{(0)}, \\
        V_{\mathrm{NLO}} & =V_{\mathrm{LO}}+V_{2 \pi}^{(2)}+V_{\mathrm{ct}}^{(2)}, \\
        V_{\mathrm{N^2LO}} & =V_{\mathrm{NLO}}+V_{2 \pi}^{(3)}, \\
        V_{\mathrm{N}^3 \mathrm{LO}} & =V_{\mathrm{N^2LO}}+V_{\mathrm{ct}}^{(4)}+V_{2 \pi,\mathrm{tl}}\\
        &+V_{2 \pi,\mathrm{rc}}+V_{2 \pi,\mathrm{fd}}+V_{2 \pi,\rm cm}.
        \end{aligned} \label{eq:potential terms}
\end{equation}
\subsection{The NN scattering}
In principle, we can 
substitute the potential $V$ into the  Blankenbeclar-Sugar (BbS)  equation to obtain the 
scattering matrix $T$, as given in Ref.~\cite{LinearIntegralEquations}. 

The Bbs equation can be transformed by defining
\begin{equation}
    \widehat{V}\left({\vec p}~', \vec{p}\right) \equiv \frac{1}{(2\pi)^3}\sqrt{\frac{M_N}{E_{p^{\prime}}}} 
    V\left({\vec p}~', \vec{p}\right) \sqrt{\frac{M_N}{E_p}}
\end{equation}
and 
\begin{equation}
    \widehat{T}\left({\vec p}~', \vec{p}\right) \equiv \frac{1}{(2\pi)^3} \sqrt{\frac{M_N}{E_{p^{\prime}}}} 
    T\left({\vec p}~', \vec{p}\right) \sqrt{\frac{M_N}{E_p}},
\end{equation}
where $M_N$ is nucleon mass and $E_{p^{\prime}}$ is nucleon energy $\sqrt{M_N^2+p'^2}$.
Actually we solve the non-relativistic Lippmann-Schwinger (LS)  equation:  
\begin{equation}
    \begin{aligned}
        \widehat{T}\left({\vec p}~', \vec{p}\right)= & \widehat{V}\left({\vec p}~', \vec{p}\right)
        +\int d^3 p^{\prime \prime} \widehat{V}\left({\vec p}~', {\vec p}~''\right) \\
        & \times \frac{M_N}{p^2-p^{\prime 2}+i \epsilon} \widehat{T}\left({\vec p}~'', \vec{p}\right).
        \end{aligned}
\end{equation}
$\widehat{V}$ and $\widehat{T}$ can be regarded as the non-relativistic potential and scattering matrix, respectively. 
By parameterizing the on-shell ($p$=$p^{\prime}$) $T$-matrix elements, we can obtain the phase shifts. In this work, $\widehat{V}$ 
and $\widehat{T}$ are denoted as $V$ and $T$.

When solving the LS equation, a cutoff on $V$ at a high momenta has to be applied
to avoid infinities in the integral. 
It is convenient to choose the regulator function as \cite{Chiraleffectivefield}
\begin{equation}
    f\left(p^{\prime}, p\right)=\exp \left[-\left(p^{\prime} / \Lambda\right)^{2 n}-(p / \Lambda)^{2 n}\right] \label{eq:regulator},
\end{equation} 
and it remains unaffected in the partial wave decomposition. For the pion-exchange component, we adopt $n=2$ for 2PE and $n=4$ 
for one-pion exchange (1PE). The choice of powers $n$ in contact terms depends on the partial wave and the order.

The numerical methods in solving the LS equation can be found 
in Ref.~\cite{Computationalnuclearphysics} and the specific formulas for the calculation of phase 
shifts are given in Ref.~\cite{HighprecisionchargedependentBonn}. Note that the Stapp convention is adopted in calculations of phase shifts~\cite{HighprecisionchargedependentBonn}.

\subsection{Local projection}

In partial wave basis, we denote the potential matrix elements $ \mathrm{i}^{L-L^{\prime}}\left\langle p^{\prime}L^{\prime} S J M|V| p L S J M\right\rangle$  by 
$V^{JS}_{L'L}\left(p^{\prime},p\right)$ and $\left\langle r^{\prime}L^{\prime} S J M|V| r L S J M\right\rangle$ by 
$V^{JS}_{L'L}\left(r^{\prime},r\right)$. The coordinate-space and momentum-space matrix elements are associated by the Fourier transform
\begin{equation}
    V^{JS}_{L^{\prime}L}\left(r^{\prime}, r\right)=\frac{2}{\pi} \int_0^{\infty} p'^2 dp' p^2 dp j_{L'}\left(p' r'\right) j_{L}
    (p r) V^{JS}_{L^{\prime}L}(p',p) \label{eq:ptor},
\end{equation}
where $j_{L'}$ and $j_L$ are spherical Bessel functions. If the matrix elements in coordinate space  can be expressed by 
\begin{equation}
V^{JS}_{L^{\prime}L}\left(r^{\prime}, r\right) = V^{JS}_{L^{\prime}L}(r) \frac{\delta\left(r - r'\right)}{r'^2},
\end{equation}
such a potential is a local potential. Otherwise the potential is nonlocal.

Up to N$^2$LO, the pion-exchange terms obtained from calculations of Feynman diagrams are local. However, the relativistic corrections and the $c_i/M_N$ terms appearing at N$^3$LO, the chosen regulator,
and the $\vec{k}$-dependent contact terms, would introduce non-localities. These non-localities prohibit direct visualizations 
and it is difficult to observe range-dependent features of the potential. In this work, 
 a local projection of the potential is applied, 
which is described in Ref.~\cite{Localprojectionslowmomentum}.

The local projection can be straightforwardly obtained by integrating 
out one of the coordinates $r'$ in the coordinate space potential, 
\begin{equation}
\begin{aligned}
\mathrm{L}\left[V\left({\vec r}~', \vec{r}\right)\right] & =\delta\left({\vec r}~'- \vec{r}\right)
\int d^3 r^{\prime \prime} V\left({\vec r}~'', \vec{r}\right) \\
& \equiv \delta\left({\vec r}~'- \vec{r}\right) \bar{V}(\vec{r}),
\end{aligned}
\end{equation}
where $\mathrm{L}$ denotes the local projection, and $\bar{V}$ is the resulted local potential.
In the partial wave basis, the equation above is written as
\begin{equation}
\bar{V}^{JS}_{L^{\prime}L}(r) =\int_0^{\infty} r^{\prime 2} d r^{\prime} 
V^{JS}_{L^{\prime}L}\left(r^{\prime}, r\right) \label{eq:localprojection}.
\end{equation}
By substituting this expression into Eq.~\eqref{eq:ptor} and integrating over the free coordinates 
analytically, we obtain
\begin{equation}
\bar{V}^{JS}_{L0}(r) = \int_0^{\infty} k^2 dk j_L(kr) V^{JS}_{L0}(k, 0)
\end{equation}
and (for $L,L'>0$)
\begin{equation}
\bar{V}^{JS}_{L^{\prime}L}(r) = N_{L'} \int_0^{\infty} dk dk' \frac{k^2}{k'} j_L(kr) V^{JS}_{L^{\prime}L}(k', k) \label{eq:ftrans},
\end{equation}
where \footnote{Compared to Ref.~\cite{Localprojectionslowmomentum}, a factor of $i^{L-L'}$
  is absorbed in the definiation of matrix elements in the momentumm space.}
\begin{equation}
N_{L'} = \frac{4}{\sqrt{\pi}} \frac{\Gamma\left(\frac{L' + 3}{2}\right)}{\Gamma\left(\frac{L'}{2}\right)} 
\end{equation} 
and $\Gamma$ denotes the gamma function. 

This approach essentially averages the non-locality, retaining the main features of the potential.  
By plotting $\bar{V}$ as a function of $r$, we can provide a visual representation 
of the potential. This allows us to study the variations and the counterbalance of  contact terms and 
 pion-exchange contributions as the cutoff increases.

\section{results and discussions}

\subsection{Potential with different cutoffs}

A high cutoff in momentum space can be roughly interpreted as a small-radius cutoff in coordinate space. As elucidated in 
Refs.~\cite{Singularpotentialslimit,ProblemRenormalizationChiral},  an attractive potential of the form
\begin{equation}
V_{\mathrm{L}}(r)=-\frac{\alpha}{2 \mu r^n}
\end{equation}
together with the repulsive centrifugal barrier $l(l+1) /\left(2 \mu r^2\right)$, compose the radial potential. For $n=2$ when $\alpha$ is 
not very small and $n \geq 3$, $V_{\mathrm{L}}(r)$ dominates at small distances in low partial waves, leading to a `singular' potential. 
The potential  in the coordinate space can be separated into a short-range part and a long-range part with a sharp cutoff $R$,
\begin{equation}
V(r)=V_{\mathrm{S}}(R) \theta(R-r)+V_{\mathrm{L}}(r) \theta(r-R).
\end{equation}
One can ensure the low-energy observables by adjusting $V_S$ when $R$ is fixed. 
 As $ R \to 0$ ($\Lambda \to \infty$), 
the parameter displays a limit-cycle behavior \cite{Singularpotentialslimit} and unphysical bound states appear.

For $NN$ potentials, similar situation occurs in low partial waves. At LO, Ref.~\cite{Renormalizationonepionexchange} demonstrates that the tensor interaction of 
1PE is `singular' and spurious bound states occur in $^3P_0$, $^3S_1-{}^3D_1$ and $^3D_2$ channels. At N$^2$LO, Ref.\cite{Improvingconvergencechiral1} illustrates 
the running behavior of parameters in  $^3P_1$ channel and indicates the emergence of a spurious bound state at $\Lambda=730$ MeV.

\begin{figure*}[]
    \includegraphics[width=\textwidth]{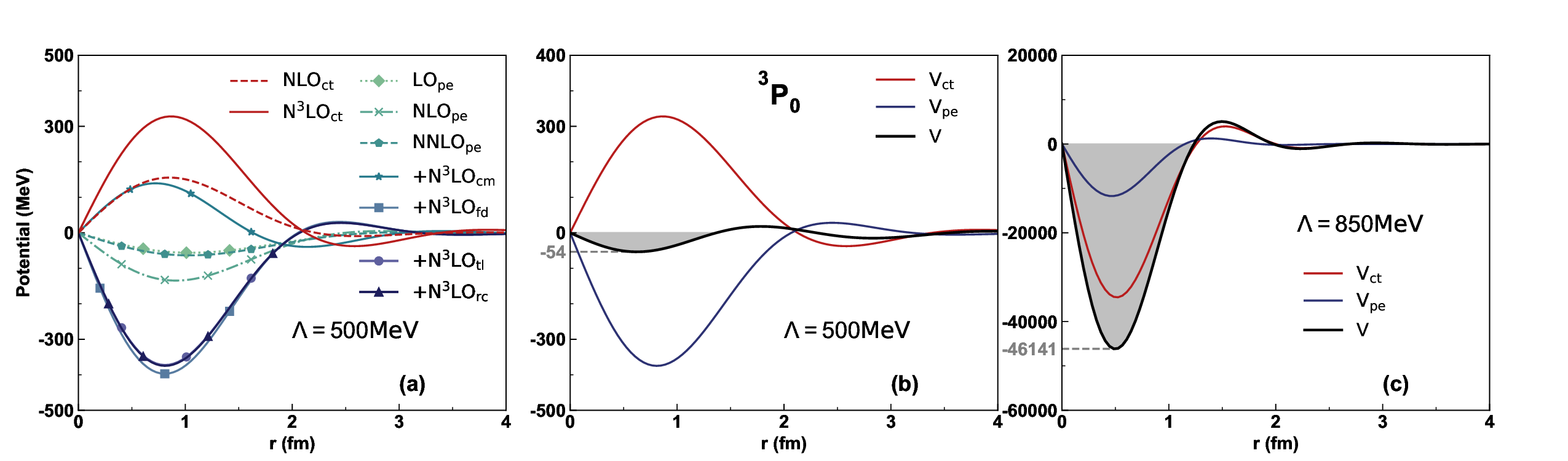}
    \caption{\label{fig:3P0} 
    Local projections of the contact  and pion-exchange components of the N$^3$LO potential in $^3P_0$ channel.
    (a) Local projections of different contributions with a cutoff of $\Lambda$=500 MeV, as described in Eq.~\eqref{eq:potential terms}.
    (b) Local projection of total nuclear potential with $\Lambda$=500 MeV.
    (c) Local projection of total nuclear potential with $\Lambda$=850 MeV, in which
  the shadow region is responsible for the appearance of deep bound states. 
    }
\end{figure*}
At N$^3$LO, based on the chiral $NN$ potential, we analyze the parameter running and the emergence of spurious bound states in partial waves with 
$J \leq 2$, as  $\Lambda$  varies from 500 MeV to 1000 MeV. In the N$^3$LO potential, there are four parameters related to  $^1S_0$ channel, and eight 
parameters related to the coupled $^3S_1-{}^3D_1$ channels. With such parameters in these channels, it is challenging to fully determine them via
phase shifts. Actually, spurious bound states can always be eliminated by adjusting the parameters in  $^1S_0$, $^3S_1-{}^3D_1$ channels at N$^3$LO below 1 GeV. 
 For $^3P_0$, $^1P_1$, $^3P_1$ channels, each have two relating parameters. Then we perform a fitting to GWU SP07 partial wave phase shifts \cite{Updatedanalysiselastic} of the $np$ channel at laboratory kinetic energy ($T_{\mathrm{lab}}$) values of 1, 5, 10, 25, 50, 100, 150, 200, 250, and 300 MeV. 
We determine the two parameters for each $P$ wave by minimizing $\chi^2=(\delta_{\mathrm{N^3LO}}-\delta_{\mathrm{GWU}})^2$ and calculate the bound states 
as the cutoff varies. We find that the first spurious bound states appear in these three $P$ waves around $\Lambda$ at 760, 800, and 850 MeV, respectively. 
Due to the sensitive dependence of the parameters on $\Lambda$ near the emergence of spurious bound states, it is difficult to precisely determine 
parameters in such  cases. For $J$=2 waves, no spurious 
bound states appear as $\Lambda$ varies from 500 to 1000 MeV. Note that, for the $^3D_2$ channel which has one related parameter, a spurious bound state appears at 
$\Lambda$ around 1100 MeV. For $\Lambda > 1200$ MeV, phase shifts of the  $^3D_2$ channel can not be described at N$^3$LO by adjusting the parameters. 
For higher partial waves,  spurious bound states 
do not appear within the considered cutoff range due to a strong centrifugal barrier.

\begin{figure}[]
    \includegraphics[width=\columnwidth]{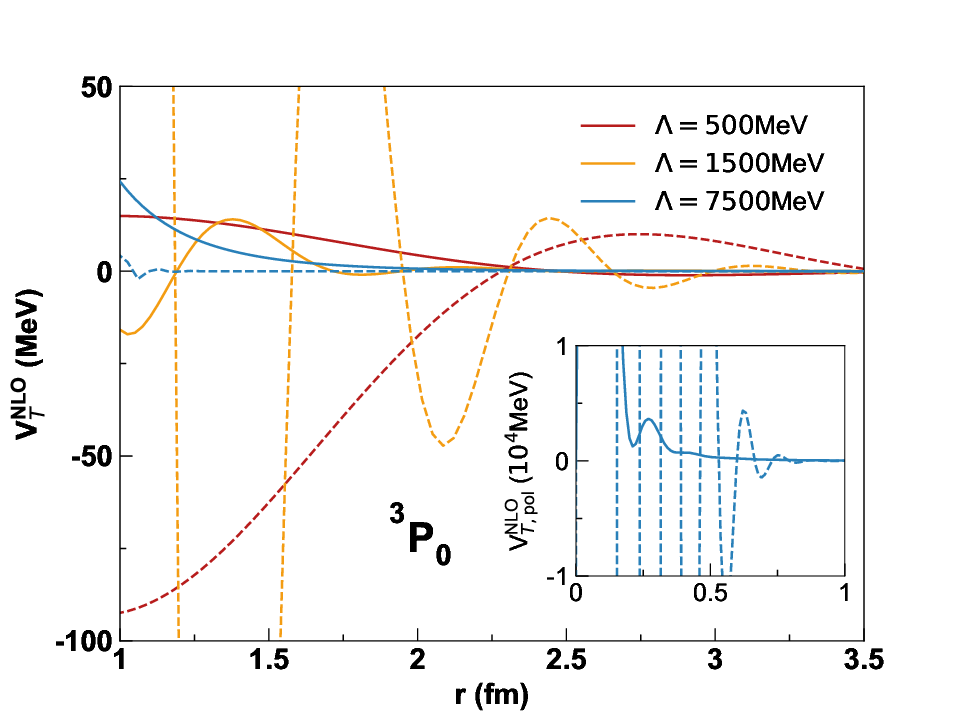} 
    \caption{\label{fig:VT_NLO}
    Local projections of contributions of the spectral function representation $V^{\mathrm{NLO}}_{T,\mathrm{SF}}$ (solid lines) and the remaining polynomial   $V^{\mathrm{NLO}}_{T,\mathrm{pol}}$ (dashed lines),
    which belong to the 2PE tensor term 
    in  $^3P_0$ channel at NLO, see the text for definitions. 
    The red, orange and blue lines correspond to local projections of potentials with cutoffs of  500 MeV, 1500 MeV, and 7500 MeV, respectively.
    In the insert, the results with $\Lambda=7500$ MeV are shown in the range from 0 to 1 fm. }  
\end{figure}
\begin{figure*}[]
    \includegraphics[width=\textwidth]{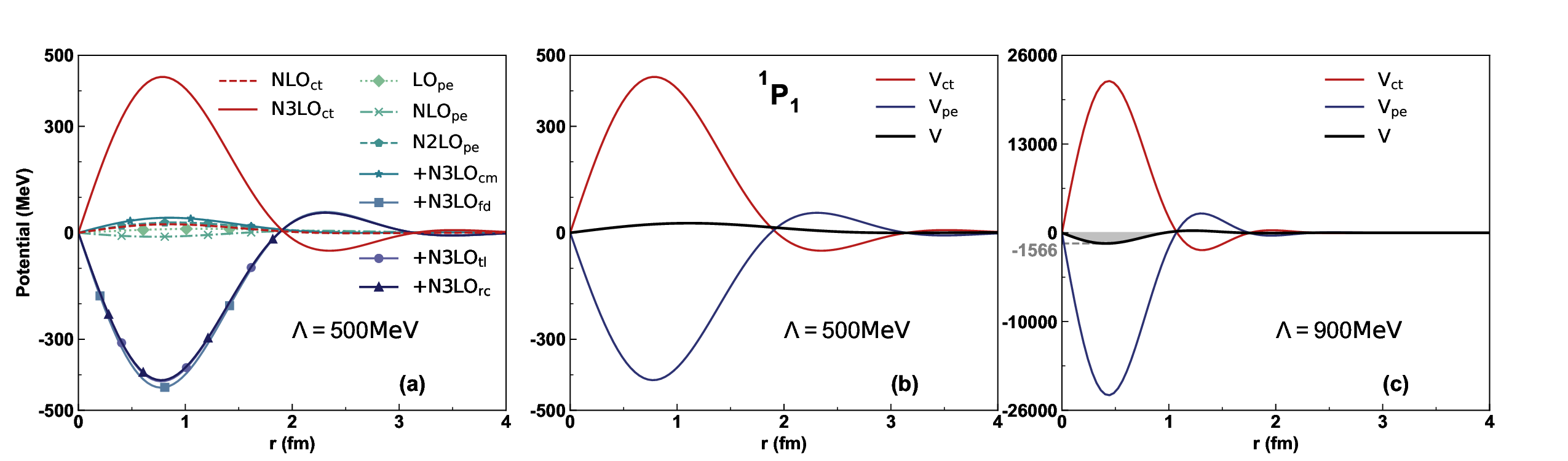}
    \caption{\label{fig:1P1}
        Local projections of the contact and pion-exchange components of the N$^3$LO potential in $^1P_0$ channel, similar to
 Fig.~\ref{fig:3P0}.}
\end{figure*}
\begin{figure*}[]
    \includegraphics[width=\textwidth]{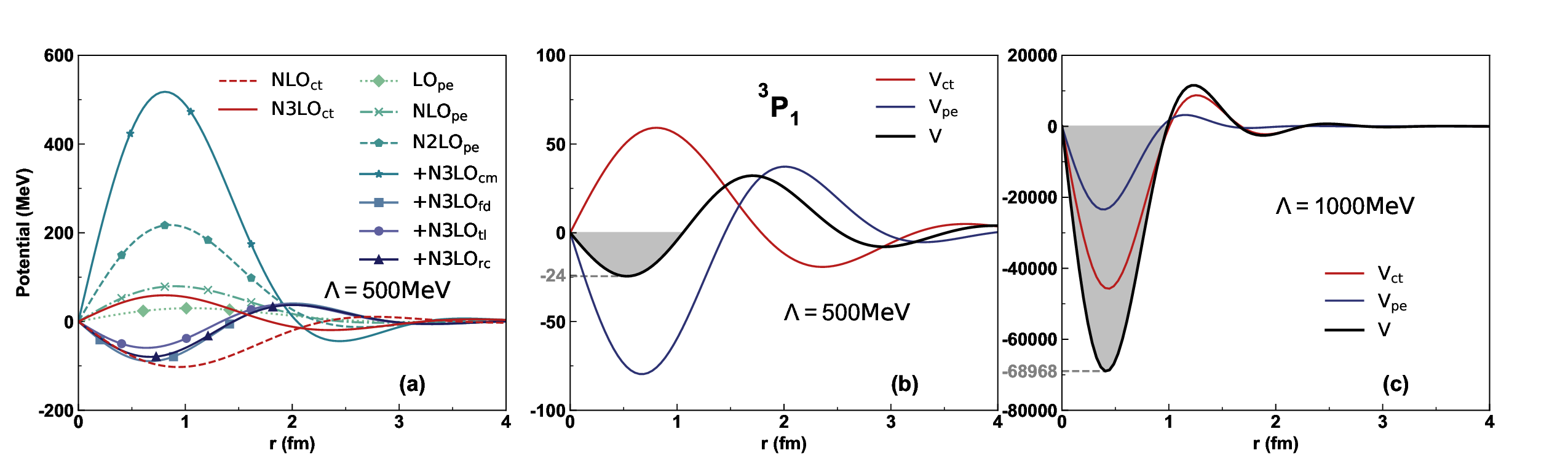}
    \caption{\label{fig:3P1}
       Local projections of the contact and pion-exchange components of the N$^3$LO potential in $^3P_1$ channel, similar to Fig.~\ref{fig:3P0}.}
\end{figure*}
\subsection{Visualization}
Based on the local projection method, we expect to observe the variations in the pion-exchange and contact-term components as the cutoff changes.
In Fig.~\ref{fig:3P0}, local projections are applied to  N$^3$LO potential \cite{Highqualitytwonucleonpotentials} as the cutoff $\Lambda$ varies in $^3P_0$ channel. 
The regulator powers related to contact terms $C_{^3P_0}$ and $D_{^3P_0}$ are $n=2$ and $n=3$. Potentials with  $\Lambda=$500 MeV and $\Lambda=$850 MeV are selected to illustrate cases without and with spurious bound states, respectively.
Note that the potential with  $\Lambda=$500 MeV shown in Fig.~\ref{fig:3P0} refers to EMN500 from Ref.~\cite{Highqualitytwonucleonpotentials}.

To verify the local projection, the bound states are calculated in both the momentum space and the projected local potential. 
At $\Lambda$ of 850 MeV for $^3P_0$, the first eigenvalues before and after local projection are 
-44964 MeV and -41180 MeV, respectively. 
At $\Lambda$ of 820 MeV, the eigenvalues before and after local projection are 
-40431 MeV and -37531 MeV.  
This indicates that the local projection includes the main features of the potential.
Note that the fitting is difficult as $\Lambda$  near the instable region.

At $\Lambda=500$ MeV, which is a safe choice for chiral nuclear forces, the pion exchange term is observed to predominantly contribute the attraction part, 
with a width of about 2 fm and a depth of approximately 400 MeV, as shown in Fig.~\ref{fig:3P0}(b). The contribution from the contact term is primarily repulsive, with a width and scale 
comparable to the pion exchange term. The contact and pion exchange terms have opposite signs and counterbalance each other. The resulting total potential 
has slight oscillations with an attraction part in the short range and a slight attraction around 2.7 fm. 
The attraction potential in the core would be overshadowed by the
strong centrifugal barrier, and has no bound-state solutions. 

In  Fig.~\ref{fig:3P0}(c), as the cutoff $\Lambda$ increases, the oscillation width of the pion exchange potential gradually decreases, while its oscillation amplitude increases. At 
$\Lambda=850$ MeV, the width is about 1.3 fm. The description of phase shift requires  that the contact term is also attractive at short range.
Note that the phase shift at this cutoff is increased by 
$\pi$ compared to that at 500 MeV.
The shadow region is related to the appearance of bound states, which is also shown but is small at $\Lambda=$500 MeV.  
In this case, the total potential is significantly attractive at short range and results in spurious deep bound states.

The decomposed potentials from different terms at N$^3$LO are displayed in Fig.1(a).
The pion-exchange contributions, contact contributions  are shown separately. 
The separated terms  are given in Eq.~\eqref{eq:potential terms}.
The oscillations in different terms up to N$^2$LO are similar at short ranges although their signs and amplitudes are different.
All $P$-wave contributions after the local projection are zero at $r=0.0$ fm due to the Bessel function in Eq.~\eqref{eq:ftrans} .
At N$^3$LO, the $c_i^2$ terms(fd) have strong attraction contribution, while the $c_i/M_N$ terms(cm) are significantly repulsive.
These two contributions are much larger than all other terms due to the large value of $c_i$. 
The contributions from the tl and rc terms are relatively small. The different terms are given in details in the Appendixes. 

If the potential in the moment space is not truncated by the regulator, the Fourier transform of the  irreducible 2PE terms is problematic \cite{Peripheralnucleonnucleonphaseb}. 
We can obtain the potential in coordinate space through its spectral-function representation \cite{epelbaum2004improving}, as given in Appendix C. 
As an example, the  tensor term of the pion exchange contribution at NLO can be divided 
into  $V^{\mathrm{NLO}}_{T,\mathrm{SF}}$ and $V^{\mathrm{NLO}}_{T,\mathrm{pol}}$, as given in Eq.~\eqref{eq:NLOVTsep}.
$V^{\mathrm{NLO}}_{T,\mathrm{SF}}$ is related to the spectral-function and converges in the Fourier transformation as the cutoff increases. 
The remaining contribution $V^{\mathrm{NLO}}_{T,\mathrm{pol}}$ is a constant at NLO in the momentum space.
Fig.~\ref{fig:VT_NLO} displays the local projections of these two terms as the cutoff $\Lambda$ varies.
The $V^{\mathrm{NLO}}_{T,\mathrm{SF}}$ term converges numerically in coordinate space for $r$$>$1 fm at a cutoff of 7500 MeV.
The regulator with a lower cutoff causes this term to oscillate around 
the converged value.
The dashed-lines denote the remaining contribution $V^{\mathrm{NLO}}_{T,\mathrm{pol}}$ with the regulator. 
With increasing cutoffs, the oscillations in both $V^{\mathrm{NLO}}_{T,\mathrm{SF}}$ and $V^{\mathrm{NLO}}_{T,\mathrm{pol}}$ become more significant.
In addition, the oscillations quickly decays at long ranges as $\Lambda$ increases. 
This is consistent with Fig.1(c).
At $\Lambda$=7500 MeV, both terms have significant oscillations only at very short ranges. 
Or say the long range part of the potential becomes convergent as $\Lambda$ increases. 
Note that the cutoff $\Lambda$ should be smaller than $\Lambda_{\chi}$ $\sim$1 GeV for physical studies. 
Generally the oscillation amplitudes in $V^{\mathrm{NLO}}_{T,\mathrm{pol}}$ is much more significant than that of $V^{\mathrm{NLO}}_{T,\mathrm{SF}}$. 
The remaining contribution is also a polynomial at N$^3$LO, which causes problems in the Fourier transform as $q$ increases. 
The remaining contributions can also be absorbed into contact terms, so that the oscillations can be much suppressed. 
In the local chiral potential, only the contribution of spectral-function representation is included \cite{Localpositionspacetwonucleon}.

Similar studies of $^1P_1$ and $^3P_1$ by the local projection method are shown in Fig.~\ref{fig:1P1} and Fig.~\ref{fig:3P1}. 
We see again that the contributions of tl and rc are relatively small.
For $^1P_1$ channel, the contact term and the pion-exchange term have the same oscillation width.
This is because the power in the regulator being adopted is the same. 
From $\Lambda$=500 to 900 MeV, the oscillation width decreases and the oscillation amplitude increases.
The situation of $^3P_1$ is similar to that of $^3P_0$.
However, the emergence of spurious states in $^1P_1$ is different from $^3P_0$ and $^3P_1$. 
At a high cutoff, the contact potential is still in an oppositive oscillation compared to the pion-exchange potential. 

\subsection{Potential at cutoff $\Lambda=700$ MeV}
\begin{figure}[]  
    \includegraphics[width=\columnwidth]{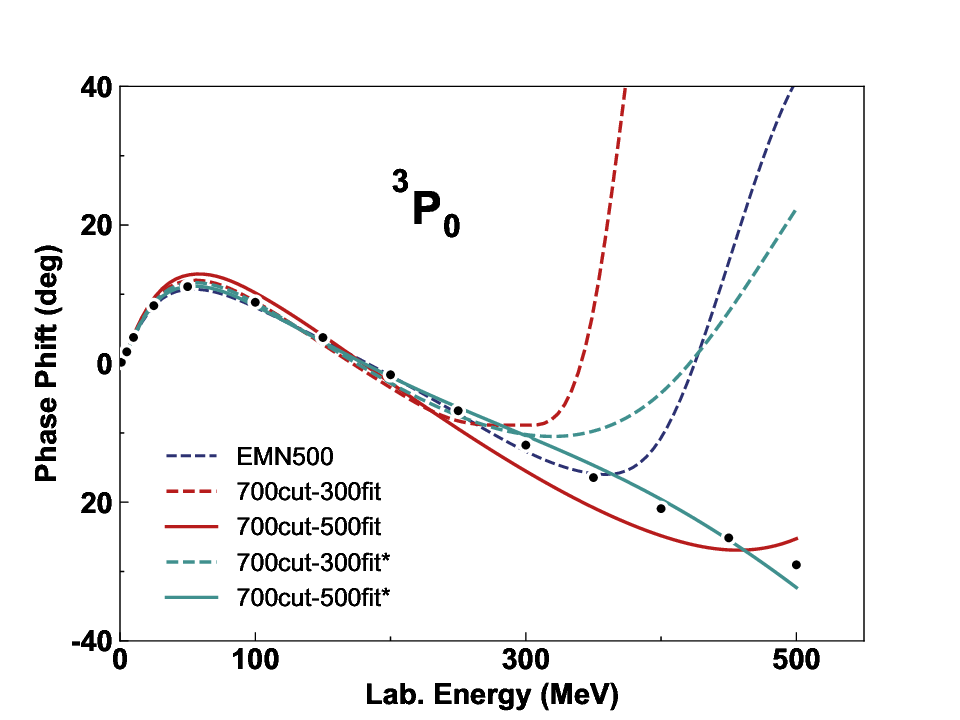} 
    \caption{\label{fig:phaseshift3P0}
    The fitting of $np$ phase shifts in $^3P_0$ channel with a cutoff of 700 MeV.
    The phase shifts up to 300 or 500 MeV are taken in the fitting  respectively.
    The red lines denote  results with $n$=2 in  regulators for both contact and pion-exchange terms.
    The green lines refer to results with different $n$ in regulators for contact terms. The circles denote
    the experimental data.  } 
\end{figure}
\begin{figure}[]  
    \includegraphics[width=\columnwidth]{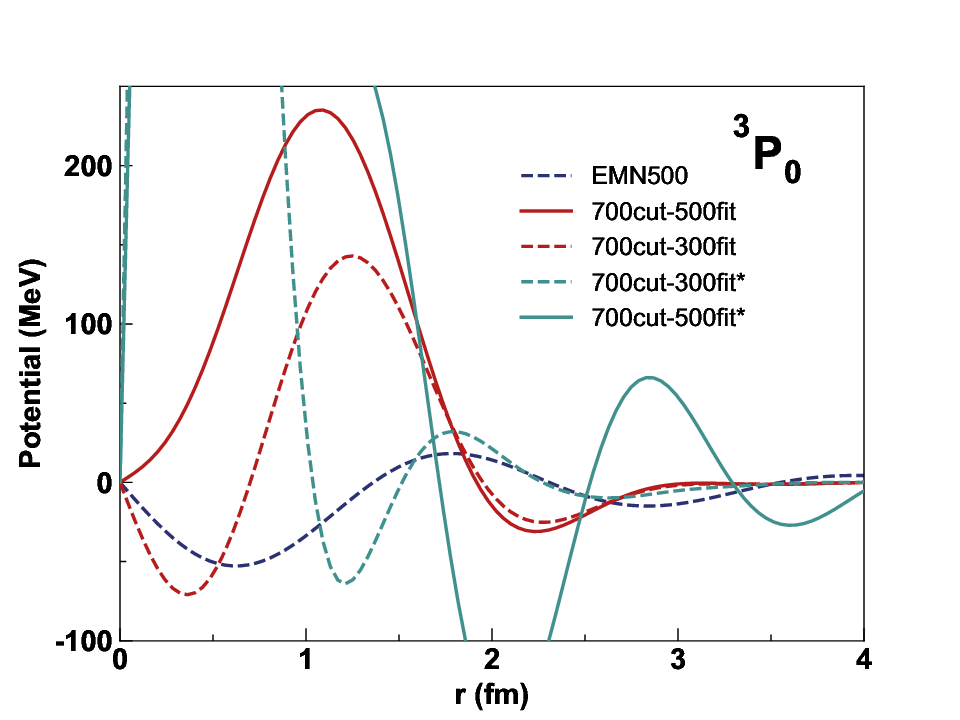} 
    \caption{\label{fig:3P0projection}
    Local projections of the potentials shown in Fig.~\ref{fig:phaseshift3P0}.}
\end{figure}

Up to date, chiral nuclear forces are usually developed with a cutoff $\Lambda$ around 500 MeV,
and the phase shifts up to 300 MeV are taken into the fitting procedure \cite{Improvedchiralnucleonnucleona,Semilocalmomentumspaceregularized, 
Accuratenuclearradii,OptimizedChiralNucleonNucleon,Accuratechargedependentnucleonnucleon,Highqualitytwonucleonpotentials}. 
Recently there are increasing strong interests to probe the short range properties of realistic nuclear forces \cite{Nucleonnucleoncorrelationsshortlived}.
For example, high energy electron scattering is a very useful tool to
study short range correlations and for the verification of the point-like nucleon models \cite{Probingcorestrong}. 
The cutoff of conventional chiral nuclear forces is not sufficiently high to describe such measurements. 
In addition, {\it ab initio} descriptions of dense nuclear matter suffers large
uncertainties above the saturation density \cite{ChiralInteractionsNexttoNexttoNexttoLeadinga}. With more observations of gravitational waves \cite{abbott2017gw170817,abbott2018gw170817},
it is of strong interests to explore nuclear forces at a higher cutoff. 
In this context, we are developing chiral $NN$ forces at N$^3$LO by adopting a cutoff at  $\Lambda=700$ MeV
and taking into account the phase shift data up to 500 MeV. 

In this work, we only adjust the $NN$ contact terms to fit the low partial wave phase shifts. 
The scattering experimental phase shifts up to 500 MeV are included. 
For $T_{\mathrm{lab}}$ energy between 0 and 10 MeV, we adopt data from the Nijmegen multienergy $np$ phase-shift analysis PWA93 \cite{Partialwaveanalysisall}.
For $T_{\mathrm{lab}}$ above 10 MeV, we adopt the phase 
shift analysis of GWU SP07.
In the range  where both sets of analysis are available, the data differences are very small.
We denote the cutoff in the phase-shift database as $T^{\mathrm{cut}}_{\mathrm{lab}}$.
The optimization is achieved by using the genetic algorithm~\cite{GA}, which is changed into an OpenMP parallel scheme in this work.

\begin{figure}[]  
    \includegraphics[width=\columnwidth]{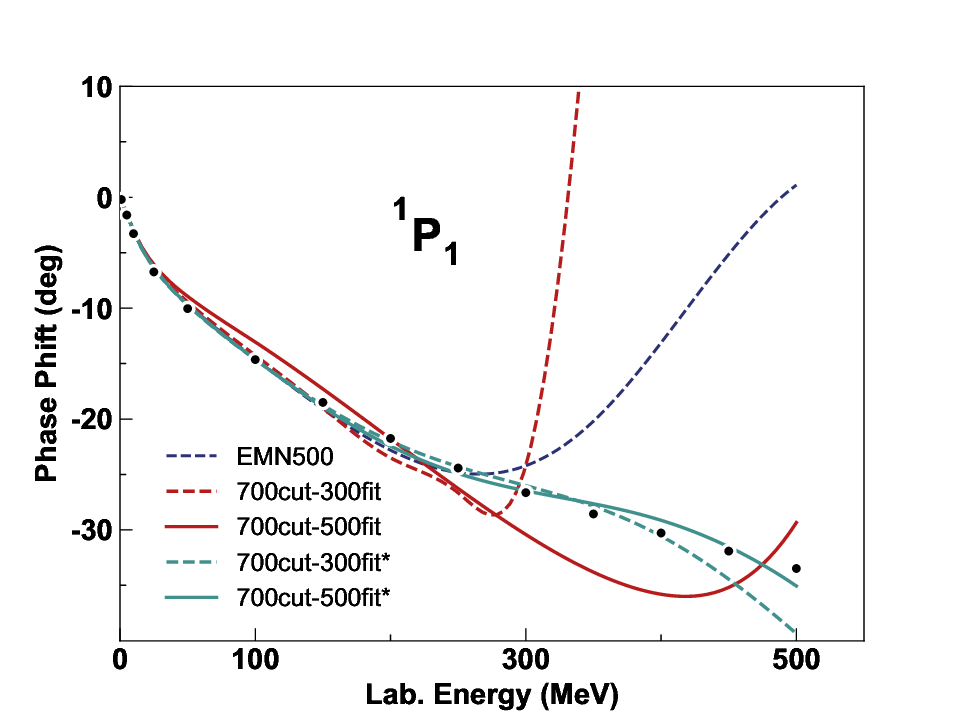} 
    \caption{\label{fig:phaseshift1P1} 
    Same as Fig.~\ref{fig:phaseshift3P0}, but for $^1P_1$ channel.}  
\end{figure}
\begin{figure}[]  
    \includegraphics[width=\columnwidth]{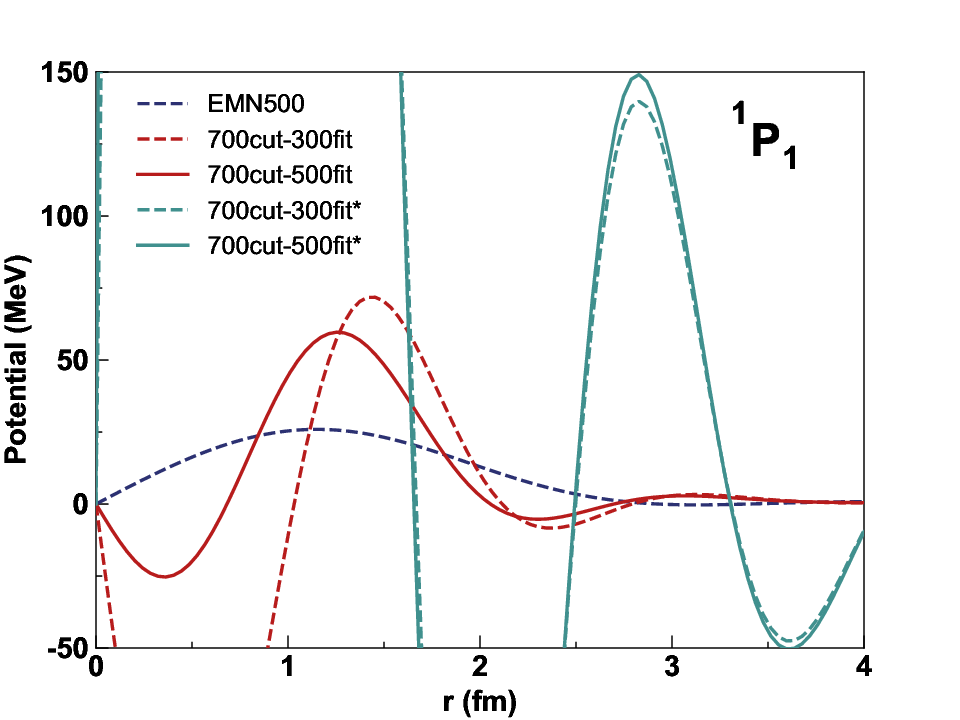} 
    \caption{\label{fig:1P1projection} 
    Same as Fig.~\ref{fig:3P0projection}, but for $^1P_1$ channel.}  
\end{figure}
\begin{figure*}
    \includegraphics[width=\textwidth]{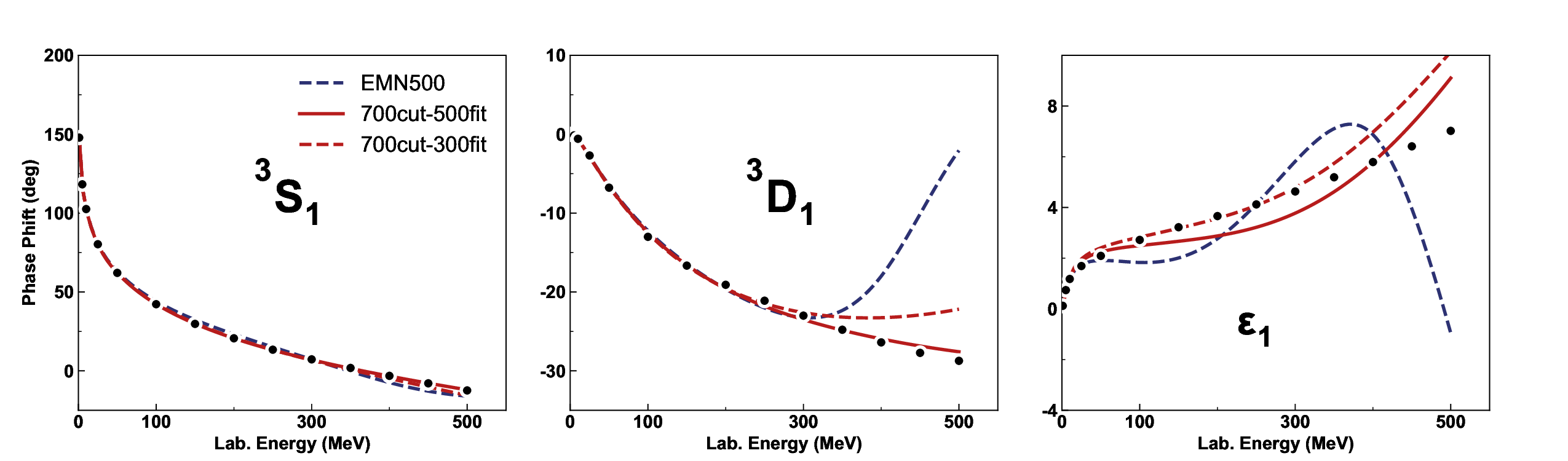}
    \caption{\label{fig:3SD1phaseshifts1}
    Same as Fig.~\ref{fig:phaseshift3P0},but for coupled ${}^3S_1-{}^3D_1$ channels.}
\end{figure*}
\begin{figure*}
    \includegraphics[width=\textwidth]{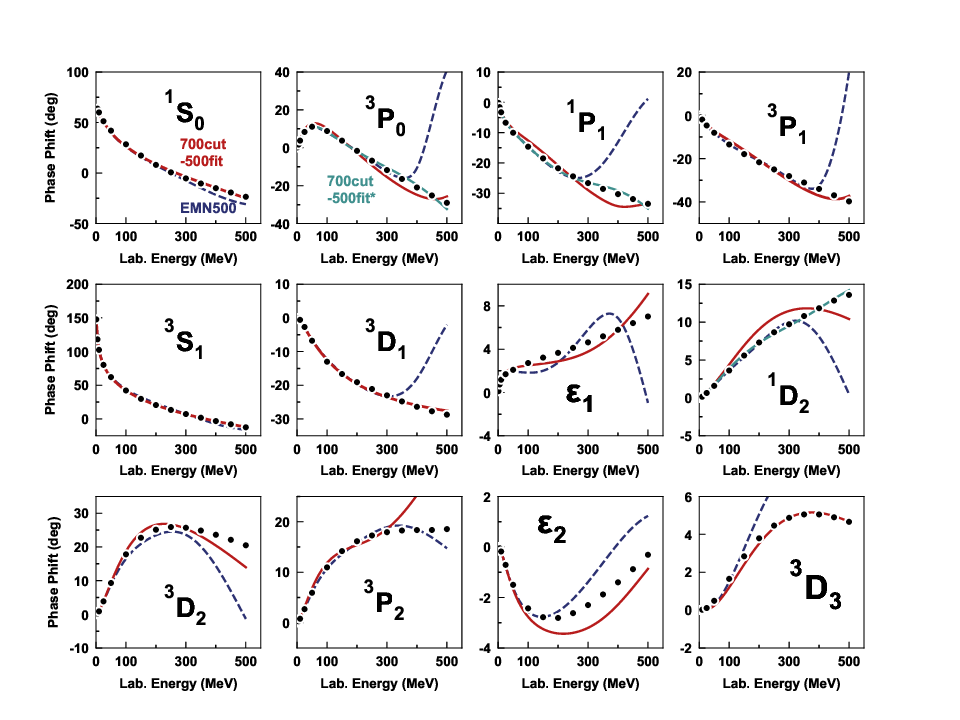}
    \caption{\label{fig:phaseshift}
  The descriptions of  $np$ phase shifts for $S$, $P$ and $D$ channels and mixing parameters $\varepsilon_1$ and 
    $\varepsilon_2$. The red solid, blue dashed lines and filled circles correspond to
    $NN$ potential with $\Lambda$=700 MeV, EMN500 and experimental data, respectively. 
   The  green solid lines are fitting with different $n$ in regulators for contact terms, as in Fig.~\ref{fig:phaseshift3P0}.
    }
\end{figure*}
\begin{figure*}
    \includegraphics[width=\textwidth]{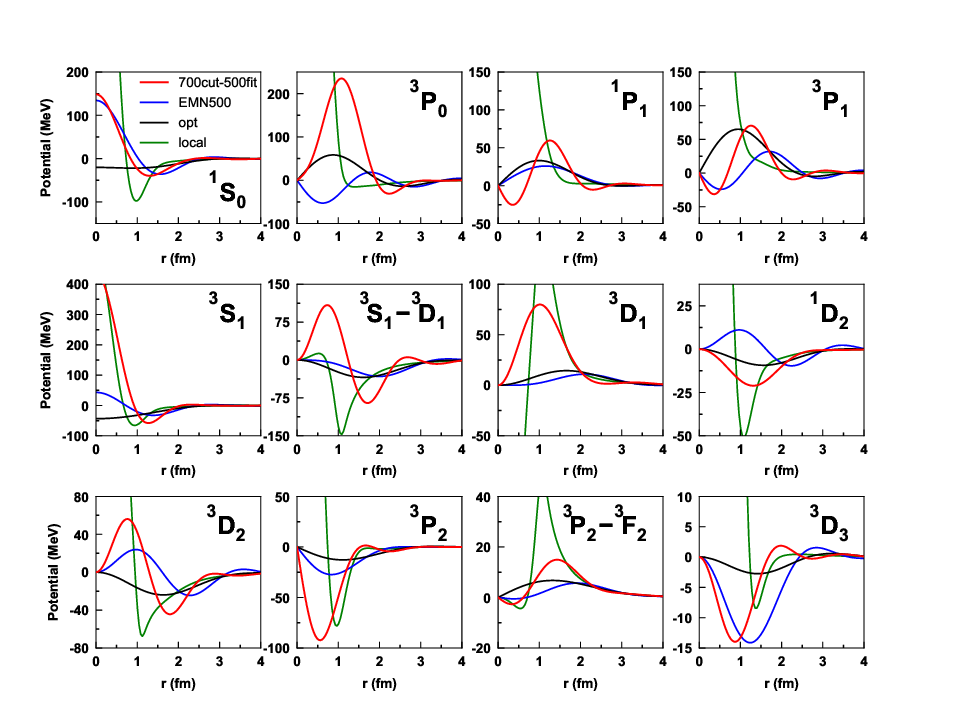}
    \caption{\label{fig:local_projection}
    Local projections of $NN$ potentials for $S$, $P$ and $D$ channels. The red lines correspond to  $NN$ potentials fitted with $\Lambda$=700 MeV.
    The  blue lines, black lines and green lines correspond to EMN500, N$^2$LO$_{\rm opt}$, and local chiral potential, respectively.}
\end{figure*}

Charge dependence in $NN$ interactions must be taken into account \cite{Chiraleffectivefield}.  
The charge dependence  due to pion-mass splitting in the one-pion exchange (1PE)
and due to nucleon-mass splitting are considered. 
Besides, the irreducible 
$\pi-\gamma$ exchange contribution in the $np$ channel is included \cite{ElectromagneticCorrectionsOnePionExchange}. 
 The charge-dependent nonderivative $\tilde{C}^{\mathrm{pp}}_{^1S_0}$,
$\tilde{C}^{\mathrm{np}}_{^1S_0}$ and $\tilde{C}^{\mathrm{nn}}_{^1S_0}$ contact terms are employed to absorb 
the remaining corrections.
And they are determined by $NN$ scattering 
lengths.

Generally, in minimizing the chi-square ($\chi^2=(w\times(\delta_{\mathrm{N^3LO}}-\delta_{\mathrm{expt}}))^2$) for fitting parameters, we adopt phase shifts at 
$T_{\mathrm{lab}}$=1, 5, 10, 25, 50, 100, 150, 200, 250, 300, 350, 400, 450, and 500 MeV. 
Due to large uncertainties in phase shifts between 300 and 500 MeV, and the influences of  inelastic scattering,
 the contribution of this part to the chi-square is multiplied by a small weight coefficient as
$w$=0.3. For $S$-waves with a strong attraction and significant phase shift values at low energies, 
additional phase shifts at $T_{\mathrm{lab}}$=0.01, 0.02,..., 0.1, 0.2,..., 1, 2,..., 10 MeV are selected in the low-energy range 
to better describe bound states and scattering lengths. We adjust charge-dependent parameters 
$\tilde{C}^{\mathrm{pp}}_{^1S_0}$ and $\tilde{C}^{\mathrm{nn}}_{^1S_0}$ to reproduce $nn$ and $pp$ scattering lengths in ${}^1S_0$. 
Note that in calculations of the $pp$ scattering length, the Coulomb interaction is taken into account appropriately via the Vincent-Phatak method \cite{Accuratemomentumspacemethoda}.

The obtained low-energy scattering parameters are shown in Table.~\ref{tab:scattering length and effective ranges}. These scattering parameters are
obtained by using the effective range expansion, as given in Ref.~\cite{HighprecisionchargedependentBonn}.
$a^C$ and $r^C$ refer to parameters by concluding Coulomb force, and $a^N$ and $r^N$ denote parameters
determined with nuclear interactions only.

\begin{table}[b]
    \caption{\label{tab:scattering length and effective ranges}
    Scattering lengths ($a$) and effective ranges ($r$) in units of fm given by the $NN$ potential with $\Lambda$=700 MeV. 
    $a_{p p}^C$ and $r_{p p}^C$ refer to  $pp$ parameters with Coulomb force. $a^N$ and $r^N$ denote parameters
    determined without Coulomb potential. The $^3S_1$ effective range parameters 
    $a_t$ and $r_t$ are also listed. }
    \begin{ruledtabular}
    \begin{tabular}{lccc}
        & $\mathrm{N}^3 \mathrm{LO}_{700}$ & \text { Empirical }& \text{References}  \\
        \colrule 
        $a_{p p}^C$ &-7.82 & -7.8196(26)& \cite{Phaseshiftanalysisb}\\
        $r_{p p}^C$ & 2.735 & 2.790(14)& \cite{Phaseshiftanalysisb}\\
        $a_{n n}^N$ & -17.1 & &\\
        $r_{n n}^N$ & 2.87 & &\\
        $a_{n n}^N$ & -18.95 & -18.95(40)& \cite{Measurementneutronneutronscattering}\\
        $r_{n n}^N$ & 2.80 & 2.75(11)& \cite{Chargesymmetryquarksa}\\
        $a_{n p}$ & -23.83 & -23.740(20)& \cite{HighprecisionchargedependentBonn} \\
        $r_{n p}$ & 2.70 & 2.77(5)& \cite{HighprecisionchargedependentBonn}\\
        $a_t$ & 5.424 & 5.419(7) & \cite{HighprecisionchargedependentBonn}\\
        $r_t$ & 1.769 & 1.753(8) & \cite{HighprecisionchargedependentBonn}\\
    \end{tabular}
    \end{ruledtabular} 
\end{table}

The $np$ scattering phase shifts in the $^3P_0$ channel are shown in Fig.~\ref{fig:phaseshift3P0}. 
We fit the parameters using phase shifts database cutoff $T^{\mathrm{cut}}_{\mathrm{lab}}$=300 and 
500 MeV with a cutoff in the regulator of $\Lambda=$ 700 MeV and $n$=2, obtaining chi-squared per datum 
values ($\chi^2$/datum) of 2.4 and 1.6, respectively. 
However, by setting $n$=1 for $C_{^3P_0}$ and $n$=2 for $D_{^3P_0}$ in contact terms, 
the fitting with  phase shifts between 0 and 300 MeV can achieve a $\chi^2$/datum of 0.54.
By setting $n$=1 for $C_{^3P_0}$ and $n$=3 for $D_{^3P_0}$,
the fitting with  phase shifts between 0 and 500 MeV can achieve a $\chi^2$/datum of 0.44.
The fitting of phase shifts can be much improved by adopting different regulators in
contact terms and pion-exchange terms. 

The local projection of the fitted potential in $^{3}P_0$ channel is shown in Fig.~\ref{fig:3P0projection}. 
We find that by adjusting $n$ the phase shift data can be well reproduced up to 500 MeV,
however, the corresponding potential exhibits strong oscillations.
There is a notable
deep attractive part around 2 fm. 
This is because of the different  powers $n$ in contact and pion-exchange terms,
the resulted oscillation widths in the coordinate space are different,
and 
the oscillations in pion-exchange terms 
can not be well absorbed by the contact terms.
In this case, although the phase shift can be
well described, the resulting potential has unrealistic oscillations in the medium range,
which is not explicit in the momentum space. 
In Ref.~\cite{Chiraleffectivefield}, some channels are also fitted with different powers $n$ in regulators for contact and pion-exchange terms,
but the oscillation is not a serious problem with a lower cutoff.

A similar situation occurs in the $^1P_1$ channel. We present the phase shifts in 
Fig.~\ref{fig:phaseshift1P1} and local projections in Fig.~\ref{fig:1P1projection}. For fittings with phase shifts 
up to $T^{\mathrm{cut}}_{\mathrm{lab}}$=300 and 500 MeV, by adopting 
$n$=1 for $C_{^1P_1}$ and $n$=3 for $D_{^1P_1}$ in regulators for contact terms, 
the phase shifts can be well described.
However, the cases with better phase shifts in Fig.~\ref{fig:phaseshift1P1} 
correspond to strong oscillations in Fig.~\ref{fig:1P1projection}.
Note that there is no spurious bound states with such oscillations. 
For fitting with data up to 300 MeV, there is a strong attraction at short ranges
and a steep-rising phase shift above 300 MeV.
In both $^1P_1$ and $^3P_1$ at high energies, the phase shifts are too repulsive
and this is also shown in the local projection around 1.2 fm.  

The phase shifts for the coupled ${}^3S_1-{}^3D_1$ channels are shown in 
Fig.~\ref{fig:3SD1phaseshifts1}. It is found that the fitting up to $T_{\mathrm{lab}}^{\rm cut}$=300 MeV in  $^3S_1$
channel extrapolates well to 500 MeV. 
In 
$^3D_1$ channel, the fitting up to 500 MeV results in a substantial improvement compared to EMN500.
 Both sets of 
parameters show improvements over EMN500 for the mixing parameter $\varepsilon_1$.
The ground state of the deuteron is a $1^+$ state, which depends solely on the coupled 
${}^3S_1-{}^3D_1$  channels. In Table~\ref{tab:deuteron}, we calculate the deuteron properties 
using the nuclear forces in Figure~\ref{fig:3SD1phaseshifts1}, with comparison with the 
EMN500 results and empirical values. The binding energy is further optimized
by adjusting the nonderivative 
$^3S_1$ term.  Compared to EMN500, a higher 
$D$-state probability $P_D$ reflects a stronger tensor force in this potential. 
This is  consistent with a larger mixing parameter $\varepsilon_1$ obtained between 50 MeV and 200 MeV  in Fig.~\ref{fig:3SD1phaseshifts1}. 
Consequently, this results in a larger asymptotic $D/S$ state $\eta$ and a larger quadrupole moment $Q_d$.

\begin{figure}[]  
    \includegraphics[width=\columnwidth]{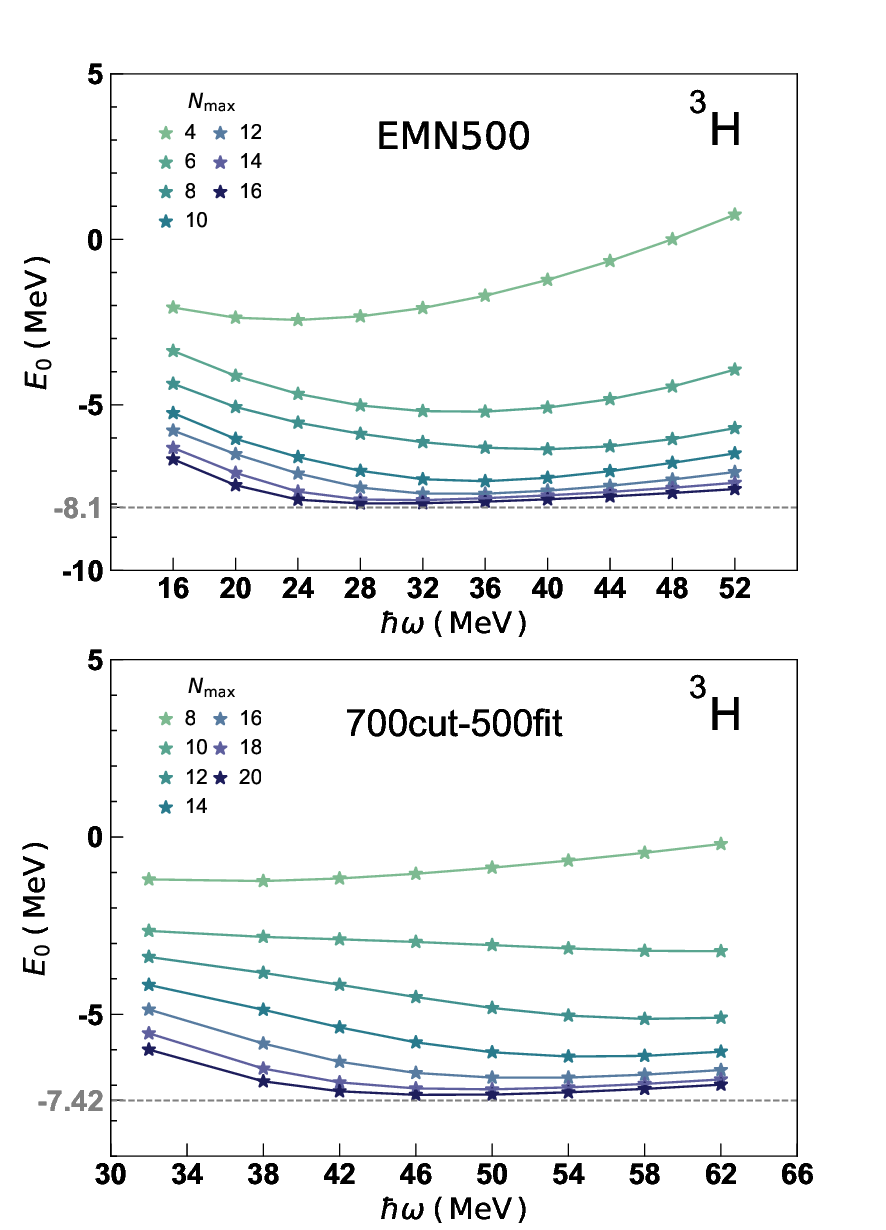} 
    \caption{\label{fig:H3}
    Calculations of binding energies of $^3\mathrm{H}$ by NCSM with different HO basis. 
    The $NN$ potentials of EMN500 and our potential with $\Lambda$=700 MeV are used.
    The convergence of results  with different basis truncations $N_{\mathrm{max}}$  as a function of
    HO frequency is shown. }   
\end{figure}

\begin{table}[b]
    \caption{Deuteron properties as predicted by $NN$ potentials at N$^3$LO with $\Lambda$=700 MeV.
    The phase shifts up to 300 or 500 MeV  are taken in the fitting respectively. 
    The properties include  binding energy $B_d$, asymptotic $S$ state $A_S$, asymptotic $D/S$ state $\eta$, 
    structure radius $r_{\mathrm{str}}$, quadrupole moment $Q_d$, $D$-state probability $P_D$.}
    \label{tab:deuteron}
    \begin{ruledtabular}
    \begin{tabular}{@{}lccccc@{}}
                        &500fit& 300fit & EMN500 & Empirical
                        \footnote{See Table XVIII of Ref.~\cite{HighprecisionchargedependentBonn} for references; 
                        the empirical value for $r_{\mathrm{str}}$ is from Ref.~\cite{Hydrogendeuteriumisotopeshift}.} \\
     \colrule
    $B_d$ (MeV)         & 2.224602 & 2.224625 & 2.224615 & 2.224575(9) \\
    $A_S$ (fm$^{-1/2}$) & 0.8879   & 0.8873   & 0.8853   & 0.8846(9) \\
    $\eta$              & 0.0260   & 0.0263   & 0.0257   & 0.0256(4) \\
    $r_{\mathrm{str}}$ (fm)          & 1.975    & 1.974    & 1.970    & 1.97507(78) \\
    $Q_d$ (fm$^2$)      & 0.2733   & 0.2768   & 0.271    & 0.2859(3) \\
    $P_D$ (\%)          & 4.35     & 4.60     & 4.15     & - \\
    \end{tabular}
\end{ruledtabular}
\end{table}

The phase shifts of $S$, $P$, and $D$ channels and mixing parameters $\varepsilon_1$ and $\varepsilon_2$ 
are displayed in red lines in Fig.~\ref{fig:phaseshift}, which are obtained with $\Lambda=700$MeV and $n=2$ in the regulators.
The phase shifts in  $^1D_2$ channel can be improved by adjusting $n$=2.5 in 
 the contact term $D_{^1D_2}$, but this would result in strong oscillations. In $^3D_2$ channel, 
the phase shifts at high energies are improved by adopting $\Lambda$=700 MeV
compared to EMN500.
For the $^3P_2$ channel, it is difficult to reproduce the phase shifts at low and high energies simultaneously.
The $^3D_3$ partial wave can be very well described with $\Lambda$=700 MeV. 
The phase shifts in some channels can be better fitted by 
by adjusting the power $n$ in the contact regulators, which are displayed in green-solid lines. 

Fig.~\ref{fig:local_projection} shows the local projections of our $NN$ potential corresponding to Fig.~\ref{fig:phaseshift}. 
For comparison, we also present the local projections of 
 EMN500,  $\mathrm{N^2LO}_{\mathrm{opt}}$ \cite{OptimizedChiralNucleonNucleon} and the local potential $\mathrm{N^3LO}_{\mathrm{Local}}$ \cite{Localpositionspacetwonucleon}.
 Here the local potential has a cutoff $R_\pi=1.0$ fm for pion-exchange terms
 and a cutoff $R_{\mathrm{ct}}=0.7$ fm for contact terms \cite{Localpositionspacetwonucleon}.
 The nuclear potentials can be different although they all reproduce the phase shifts.
As we have discussed, the nonlocal $NN$ potential exhibits oscillatory behavior at short ranges.
It is difficult to separate the short-range and  long-range contributions distinctly. 
In contrast, there is a distinct separation between long-range 
and contact contributions in local potentials. 
Nonlocal $NN$ potentials are generally 
softer compared to the local potential. 
Our potential with $\Lambda=700$ MeV is harder compared to EMN500.
It can be seen that the $\mathrm{N^2LO}_{\mathrm{opt}}$ nuclear force is the softest amongst them.

In Fig.~\ref{fig:H3}, we show ground-state energy of  $^3\mathrm{H}$, which is obtained in {\it ab initio} no core shell model (NCSM) \cite{initionocore}.
The NCSM code is taken from Ref.~\cite{Michel2021}. 
In NCSM calculations,  results are dependent on the frequency $\hbar \omega$ of the harmonic oscillator (HO) basis
and also the truncation parameter $N_{\mathrm{max}}$ of the model space. 
The NCSM calculations with our potential (${\rm N^3LO}_{700}$ with $\Lambda$=700 MeV) converge slower
than that with EMN500.
To obtain convergence with an uncertainty of 1$\%$, ${\rm N^3LO}_{700}$  needs $N_{\mathrm{max}}$=24,
while EMN500 needs only $N_{\mathrm{max}}$=20.
The optimal HO frequency in calculations with
${\rm N^3LO}_{700}$ (around $\hbar \omega$=48 MeV) is much larger than the that with EMN500 (around $\hbar \omega$=28 MeV).
Such differences between ${\rm N^3LO}_{700}$  and EMN500 are because ${\rm N^3LO}_{700}$ has a higher cutoff. 
The binding
energy with our potential is 7.42 MeV, which is less bound compared to EMN500.
The experimental binding energy of  $^3\mathrm{H}$ is 8.48 MeV.   
In this case, the role of three-body force is important. 
For $^3P_0$, $^1P_1$, and $^1D_2$ channels, if the
 potentials are replaced by
the better fitting of phase shifts in Fig.~\ref{fig:phaseshift}, the resulted binding  energy of $^3\mathrm{H}$ would be 7.50 MeV. 
To be more specific, the smaller binding energy by ${\rm N^3LO}_{700}$ is mainly due to the strong repulsive potential in
  $^3P_0$ and $^3S_1-^3D_1$.
This is because the phase shifts are fitted up to 500 MeV, which leads to a strong repulsive feature.   
If the phase shifts are fitted to  300 MeV for $^3P_0$ and $^3S_1-^3D_1$,
the obtained binding energy of $^3\mathrm{H}$ would be 7.78 MeV. 
It seems that the better descriptions of  phase shifts at high energies, the binding energy of $^3\mathrm{H}$ would be underestimated using only $NN$ potential.

\subsection{Discussions}
It has been discussed that the renormalization of chiral $NN$ interaction has been an issue \cite{ProblemRenormalizationChiral,Renormalizationchiraltwopiona}. 
The chiral $NN$ interactions has dependence on the cutoff, as also demonstrated in this work. 
The short-range properties are very much related to the regulator function. 
Regarding experiments on short-range correlation (SRC), it seems that nuclear interactions with a high cutoff are needed \cite{Nucleonnucleoncorrelationsshortlived,Probingcorestrong}. 
This presents a challenge for the conventional framework of chiral $NN$ interaction. 
For the non-local  $NN$ interaction in the momentum space, there are strong oscillations
in the coordinate space.  Such strong oscillations are more significant as the cutoff increases.
The development of local chiral potential \cite{Localchiraleffective,Minimallynonlocalnucleonnucleon,Localchiralpotentials,Localpositionspacetwonucleon} can avoid such oscillation issues. 
In our calculations, the descriptions of $P$ partial waves are not perfect as that of $S$-wave.
Some $D$ waves can also be well reproduced. 
There are unphysical deep bound states in $P$ channels as the cutoff increases.
There may be some physics missing in the current framework. 
It has been argued that the worse descriptions of $P$ waves can result in better descriptions of nuclear structures \cite{Nucleonnucleonpotentialsfull}. 
And the inclusion of $\Delta$ in chiral interaction  at N$^2$LO doesn't lead to significant differences in calculations of nuclear matter \cite{Nucleonnucleonpotentialsfull}. 
Besides,  it has been proposed the modified power counting with the perturbative solving  LS equation
to resolve the renormalization issue \cite{Renormalizingchiralnucleara,Shortrangenuclearforces}. 
The internal structures of nucleons also play a role in SRC and European
Muon Collaboration(EMC) effects \cite{Nucleonnucleoncorrelationsshortlived}, impacting our understandings of
nuclear forces and nuclear structures.

\section{Summery}

To summarize, we have investigated the properties of chiral $NN$ interactions at N$^3$LO as the cutoff increases.
The cutoff $\Lambda$ is conventionally taken as around 500 MeV.
However, hard $NN$ interactions with a high cutoff are needed to address recent experiments on SRC.
In this work, we employ the local projection method to study the range-dependent features of  $NN$ interactions in
an intuitively manner. 
With increasing $\Lambda$,  non-local $NN$ interactions display strong oscillations in coordinate space. 
The oscillation width decreases and oscillation amplitude increases as $\Lambda$ increases. 
The strong oscillations are related to the appearance of unphysical bound states in $P$ channels. 
The local projections of different pion-exchange terms are analyzed in details. 
The remaining polynomial terms in 2PE contributions after the spectral function representation
cause significant oscillations. 
There are no such unphysical bound states in $S$ and $D$ waves at N$^3$LO with cutoffs below 1 GeV. 

We also fit a chiral $NN$ interaction at N$^3$LO with a cutoff of 700 MeV. 
The phase shifts up to 500 MeV have been adopted. 
As a result, the phase shifts of $S$ and $D$ waves can be well reproduced, except for $^1D_2$ channel.
However, the fitting of  $P$ waves can not describe phase shifts at low and high energies simultaneously.  
The fitting can be improved by choosing different power $n$ in regulators for contact terms.
In this way the better fitting of phase shifts of $P$ waves would lead to unrealistic strong oscillations. 
The $NN$ interaction with 700 MeV cutoff has been testified by reproducing the properties deuterium with
a slightly larger component of $D$ waves compared to ENM500.
For $^3\mathrm{H}$, the binding energy is calculated to be 7.42 MeV, which is related to
the repulsive potential in $^3P_0$ and $^3S_1-^3D_1$. 
Actually the binding energy of $^3\mathrm{H}$ becomes to be 7.78 MeV if these two channels are fitted to 300 MeV.
We also discussed relevant issues in renormalization and $P$ waves in the framework of chiral $NN$ interactions. 

\acknowledgments
We thank for many helpful discussions with R. Machleidt and D. R. Entem, and also for helpful comments by Shuang Zhang.   
This work was supported by the National Key R$\&$D Program of China (Grant No. 2023YFE0101500,2023YFA1606403)
 and National Natural Science Foundation of China under Grants
No.  12335007.
We also acknowledge the funding support from the State Key Laboratory of Nuclear Physics and Technology, Peking University (No. NPT2023ZX01).

\appendix
\section{Pion-exchange component}
The NN potential is governed by the customary symmetries, hence it manifests in the general 
forms for a two-nucleon potential \cite{Nuclearforcesmomentum}
\begin{equation}
    \begin{aligned}
        V\left({\vec p}~', \vec{p}\right)= & V_C+\boldsymbol{\tau}_1 \cdot \boldsymbol{\tau}_2 
        W_C+\left[V_S+\boldsymbol{\tau}_1 \cdot \boldsymbol{\tau}_2 W_S\right] \vec{\sigma}_1 \cdot \vec{\sigma}_2 \\
        & +\left[V_{L S}+\boldsymbol{\tau}_1 \cdot \boldsymbol{\tau}_2 W_{L S}\right][-i \vec{S} 
        \cdot(\vec{q} \times \vec{k})] \\
        & +\left[V_T+\boldsymbol{\tau}_1 \cdot \boldsymbol{\tau}_2 W_T\right] \vec{\sigma}_1 \cdot \vec{q} \vec{\sigma}_2 
        \cdot \vec{q} \\
        & +\left[V_{\sigma L}+\boldsymbol{\tau}_1 \cdot \boldsymbol{\tau}_2 W_{\sigma L}\right] \vec{\sigma}_1 
        \cdot(\vec{q} \times \vec{k}) \vec{\sigma}_2 \cdot(\vec{q} \times \vec{k}),\label{eq:general potential}
        \end{aligned}
\end{equation}
where $\vec{S}=(\vec{\sigma}_1+\vec{\sigma}_2)/2$ is the total spin, with $\vec{\sigma}_1$, $\vec{\sigma}_2$ and 
$\boldsymbol{\tau}_1 \cdot \boldsymbol{\tau}_2$ being the spin and isospin operators. 
\subsection{Leading order}
At LO, there is only the 1PE contribution. The charge-independent 1PE is 
\begin{equation}
    V_{1 \pi}^{(\mathrm{CI})}\left({\vec p}~', \vec{p}\right)=-\frac{g_A^2}{4 f_\pi^2} \boldsymbol{\tau}_1 \cdot 
    \boldsymbol{\tau}_2 \frac{\vec{\sigma}_1 \cdot \vec{q} \vec{\sigma}_2 \cdot \vec{q}}{q^2+m_\pi^2} \label{eq:1PE}
\end{equation}
Note that, for on-shell and in the CMS, there are no relativistic corrections at any order. As discussed in Ref.~\cite{Chiraleffectivefield}, 
by adjusting the value of $g_A$, the familiar expression for 1PE, Eq.~\eqref{eq:1PE} is appropriate at least to fourth order. 
Therefore, in this paper, 1PE is represented solely by this term. 
\subsection{Next-to-leading order}
The $NN$ diagrams that occur at NLO contribute in the following way 
\begin{equation}
    \begin{aligned}
        W_C= & \frac{L(\tilde{\Lambda} ; q)}{384 \pi^2 f_\pi^4}\left[4 m_\pi^2\left(1+4 g_A^2-5 g_A^4\right)\right. \\
        & \left.+q^2\left(1+10 g_A^2-23 g_A^4\right)-\frac{48 g_A^4 m_\pi^4}{w^2}\right],
        \end{aligned}
\end{equation}
\begin{equation}
    V_T=-\frac{1}{q^2} V_S=-\frac{3 g_A^4}{64 \pi^2 f_\pi^4} L(\tilde{\Lambda} ; q). \label{eq:NLOVT}
\end{equation}
The loop function $L(\tilde{\Lambda} ; q)$ is regularized by spectral-function cutoff $\tilde{\Lambda}=650$ MeV 
\cite{Peripheralnucleonnucleonphaseb,epelbaum2004improving}
\begin{equation}
    L(\tilde{\Lambda} ; q)=\frac{w}{2 q} \ln \frac{\tilde{\Lambda}^2\omega^2+q^2s^2+2\tilde{\Lambda} q w s}{4 m_\pi^2\left(\tilde{\Lambda}^2+q^2\right)}
\end{equation}
with $w=\sqrt{4 m_\pi^2+q^2}$ and $s=\sqrt{\tilde{\Lambda}^2-4 m_\pi^2}$.
\subsection{Next-to-next-to-leading order}
The N$^2$LO contribution is given by
\begin{equation}
    V_{C}=\frac{3 g_{A}^{2}}{16 \pi f_{\pi}^{4}}\left[2 m_{\pi}^{2}\left(c_{3}-2 c_{1}\right)+c_{3} q^{2}\right]\left(2 m_{\pi}^{2}+q^{2}\right) 
    A(\tilde{\Lambda} ; q),
\end{equation}
\begin{equation}
    W_{T}=-\frac{1}{q^{2}} W_{S}=-\frac{g_{A}^{2}}{32 \pi f_{\pi}^{4}} c_{4} w^{2} A(\tilde{\Lambda} ; q).
\end{equation}
The loop function $A(\tilde{\Lambda} ; q)$ is regularized in the same way as $L(\tilde{\Lambda} ; q)$
\begin{equation}
    A(\tilde{\Lambda} ; q)=\frac{1}{2 q} \arctan \frac{q\left(\tilde{\Lambda}-2 m_{\pi}\right)}{q^{2}+2 \tilde{\Lambda} m_{\pi}} 
\end{equation}
\begin{widetext}
\subsection{Next-to-next-to-next-to-leading order}
\subsubsection{Football diagram at N$^3$LO}
The football diagram at N$^3$LO generates
\begin{equation}
    V_{C}=\frac{3}{16 \pi^{2} f_{\pi}^{4}}\left\{\left[\frac{c_{2}}{6} w^{2}+c_{3}\left(2 m_{\pi}^{2}+q^{2}\right)-4 c_{1} 
    m_{\pi}^{2}\right]^{2}+\frac{c_{2}^{2}}{45} w^{4}\right\} L(\tilde{\Lambda} ; q),
\end{equation}
\begin{equation}
    W_{T}=-\frac{1}{q^{2}} W_{S}=\frac{c_{4}^{2}}{96 \pi^{2} f_{\pi}^{4}} w^{2} L(\tilde{\Lambda} ; q).
\end{equation}
\subsubsection{$c_i/M_N$ contributions}
This contribution consists of diagrams with one vertex proportional to $c_i$ and one $1/M_N$ correction.
\begin{equation}
    V_{C}= -\frac{g_{A}^{2} L(\tilde{\Lambda}, q)}{32 \pi^{2} M_{N} f_{\pi}^{4}}\left[\left(c_{2}-6 c_{3}\right) 
    q^{4}+4\left(6 c_{1}+c_{2}-3 c_{3}\right) q^{2} m_{\pi}^{2}+6\left(c_{2}-2 c_{3}\right) m_{\pi}^{4}+24\left(2 c_{1}+c_{3}\right) m_{\pi}^{6} w^{-2}\right],
\end{equation}
\begin{equation}
 W_{C}=-\frac{c_{4} q^{2} L(\tilde{\Lambda},q)}{192 \pi^{2} M_{N} f_{\pi}^{4}}\left[g_{A}^{2}\left(8 m_{\pi}^{2}+5 q^{2}\right)+w^{2}\right], 
\end{equation}
\begin{equation}
 W_{T}=-\frac{1}{q^{2}} W_{S}=-\frac{c_{4} L(\tilde{\Lambda},q)}{192 \pi^{2} M_{N} f_{\pi}^{4}}\left[g_{A}^{2}\left(16 m_{\pi}^{2}+7 q^{2}\right)-w^{2}\right], 
\end{equation}
\begin{equation}
 V_{L S}=\frac{c_{2} g_{A}^{2}}{8 \pi^{2} M_{N} f_{\pi}^{4}} w^{2} L(\tilde{\Lambda},q), 
\end{equation}
\begin{equation}
 W_{L S}=-\frac{c_{4} L(\tilde{\Lambda},q)}{48 \pi^{2} M_{N} f_{\pi}^{4}}\left[g_{A}^{2}\left(8 m_{\pi}^{2}+5 q^{2}\right)+w^{2}\right].
\end{equation}
\subsubsection{Leading relativistic corrections}
The relativistic corrections of the NLO diagrams, count as N$^3$LO and are given by
\begin{equation}
    V_{C}=\frac{3 g_{A}^{4}}{128 \pi f_{\pi}^{4} M_{N}}\left[\frac{m_{\pi}^{5}}{2 w^{2}}+\left(2 m_{\pi}^{2}+
    q^{2}\right)\left(q^{2}-m_{\pi}^{2}\right) A(\tilde{\Lambda} ; q)\right],
\end{equation}
\begin{equation}
    W_{C}=\frac{g_{A}^{2}}{64 \pi f_{\pi}^{4} M_{N}}\left\{\frac{3 g_{A}^{2} m_{\pi}^{5}}{2 \omega^{2}}
    +\left[g_{A}^{2}\left(3 m_{\pi}^{2}+2 q^{2}\right)-2 m_{\pi}^{2}-q^{2}\right]\left(2 m_{\pi}^{2}
    +q^{2}\right) A(\tilde{\Lambda} ; q)\right\}, 
\end{equation}
\begin{equation}
    V_{T}=-\frac{1}{q^{2}} V_{S}=\frac{3 g_{A}^{4}}{256 \pi f_{\pi}^{4} M_{N}}
    \left(5 m_{\pi}^{2}+2 q^{2}\right) A(\tilde{\Lambda} ; q), 
\end{equation}
\begin{equation}
    W_{T}=-\frac{1}{q^{2}} W_{S}=\frac{g_{A}^{2}}{128 \pi f_{\pi}^{4} M_{N}}\left[g_{A}^{2}
    \left(3 m_{\pi}^{2}+q^{2}\right)-w^{2}\right] A(\tilde{\Lambda} ; q), 
\end{equation}
\begin{equation}
    V_{L S}=\frac{3 g_{A}^{4}}{32 \pi f_{\pi}^{4} M_{N}}\left(2 m_{\pi}^{2}+q^{2}\right) A(\tilde{\Lambda} ; q), 
\end{equation}
\begin{equation}
    W_{L S}=\frac{g_{A}^{2}\left(1-g_{A}^{2}\right)}{32 \pi f_{\pi}^{4} M_{N}} w^{2} A(\tilde{\Lambda} ; q). 
\end{equation}
\subsubsection{Leading two-loop contributions}
The leading-order 2PE two-loop diagrams are of N$^3$LO. In terms of spectral functions, the results are 
\begin{equation}
    \operatorname{Im} V_{C}=\frac{3 g_{A}^{4}\left(2 m_{\pi}^{2}-\mu^{2}\right)}{\pi \mu\left(4 f_{\pi}\right)^{6}}
    \left[\left(m_{\pi}^{2}-2 \mu^{2}\right)\left(2 m_{\pi}+\frac{2 m_{\pi}^{2}-\mu^{2}}{2 \mu} \ln \frac{\mu+2 m_{\pi}}
    {\mu-2 m_{\pi}}\right)+4 g_{A}^{2} m_{\pi}\left(2 m_{\pi}^{2}-\mu^{2}\right)\right], 
\end{equation}
\begin{equation}
    \begin{aligned} \operatorname{Im} W_{C}= & \frac{2 \kappa}{3 \mu\left(8 \pi f_{\pi}^{2}\right)^{3}} \int_{0}^{1} 
        d x\left[g_{A}^{2}\left(\mu^{2}-2 m_{\pi}^{2}\right)+2\left(1-g_{A}^{2}\right) \kappa^{2} x^{2}\right]\\
        &\left\{96 \pi^{2} f_{\pi}^{2}\left[\left(2 m_{\pi}^{2}-\mu^{2}\right)\left(\bar{d}_{1}+\bar{d}_{2}\right)
        -2 \kappa^{2} x^{2} \bar{d}_{3}+4 m_{\pi}^{2} \bar{d}_{5}\right]\right. \\ 
        & +\left[4 m_{\pi}^{2}\left(1+2 g_{A}^{2}\right)-\mu^{2}\left(1+5 g_{A}^{2}\right)\right] \frac{\kappa}{\mu} 
        \ln \frac{\mu+2 \kappa}{2 m_{\pi}}+\frac{\mu^{2}}{12}\left(5+13 g_{A}^{2}\right)-2 m_{\pi}^{2}\left(1+2 g_{A}^{2}\right) \\ 
        & -3 \kappa^{2} x^{2}+6 \kappa x \sqrt{m_{\pi}^{2}+\kappa^{2} x^{2}} \ln \frac{\kappa x+\sqrt{m_{\pi}^{2}+\kappa^{2} x^{2}}}{m_{\pi}} \\ 
        & \left.+g_{A}^{4}\left(\mu^{2}-2 \kappa^{2} x^{2}-2 m_{\pi}^{2}\right)\left[\frac{5}{6}+\frac{m_{\pi}^{2}}{\kappa^{2} x^{2}}
        -\left(1+\frac{m_{\pi}^{2}}{\kappa^{2} x^{2}}\right)^{3 / 2} \ln \frac{\kappa x+\sqrt{m_{\pi}^{2}+\kappa^{2} x^{2}}}{m_{\pi}}
        \right]\right\},
    \end{aligned} 
\end{equation}
\begin{equation}
    \begin{aligned}
        \operatorname{Im} V_{S}&=\mu^{2} \operatorname{Im} V_{T}=\frac{g_{A}^{2} \mu \kappa^{3}}{8 \pi f_{\pi}^{4}}
        \left(\bar{d}_{15}-\bar{d}_{14}\right)
        \\&+\frac{2 g_{A}^{6} \mu \kappa^{3}}{\left(8 \pi f_{\pi}^{2}\right)^{3}} \int_{0}^{1} d x\left(1-x^{2}\right)
        \left[\frac{1}{6}-\frac{m_{\pi}^{2}}{\kappa^{2} x^{2}}+\left(1+\frac{m_{\pi}^{2}}{\kappa^{2} x^{2}}\right)^{3 / 2} 
        \ln \frac{\kappa x+\sqrt{m_{\pi}^{2}+\kappa^{2} x^{2}}}{m_{\pi}}\right], 
        \end{aligned}
\end{equation}
\begin{equation}
    \operatorname{Im} W_{S}=\mu^{2} \operatorname{Im} W_{T}(i \mu)=\frac{g_{A}^{4}\left(4 m_{\pi}^{2}-\mu^{2}\right)}
    {\pi\left(4 f_{\pi}\right)^{6}}\left[\left(m_{\pi}^{2}-\frac{\mu^{2}}{4}\right) \ln \frac{\mu+2 m_{\pi}}{\mu-2 m_{\pi}}
    +\left(1+2 g_{A}^{2}\right) \mu m_{\pi}\right], 
\end{equation}
where $\kappa=\sqrt{\mu^2/4-m_\pi^2}$.

The potentials in (subtracted) spectral representation are given in momentum space by
\begin{equation}
    V_{C, S}(q) =-\frac{2 q^{6}}{\pi} \int_{2 m_{\pi}}^{\tilde{\Lambda}} d \mu \frac{\operatorname{Im} V_{C, S}(i \mu)}
    {\mu^{5}\left(\mu^{2}+q^{2}\right)},  
\end{equation}
\begin{equation}
    V_{T, L S}(q)  =\frac{2 q^{4}}{\pi} \int_{2 m_{\pi}}^{\tilde{\Lambda}} d \mu \frac{\operatorname{Im} V_{T, L S}(i \mu)}
    {\mu^{3}\left(\mu^{2}+q^{2}\right)}. \label{eq:SFR}
\end{equation}
\end{widetext}
\section{Contact terms}
The contact terms are given in the general form Eq.~\eqref{eq:general potential}, with $V_\alpha$ and $W_\alpha$ 
$(\alpha=C,S,LS,T,\sigma L)$ adopting polynomials in terms of $\vec{k}^2$, $\vec{q}^2$, and $\left(\vec{k}\times\vec{q}\right)^2$. 
Terms that include a factor $\boldsymbol{\tau}_1 \cdot \boldsymbol{\tau}_2$ can be left out due to Fierz ambiguity.

To be specific, the LO contact potential is given by
\begin{equation}
    V_{\mathrm{ct}}^{(0)}\left({\vec p}~', \vec{p}\right)=C_S+C_T \vec{\sigma}_1 \cdot \vec{\sigma}_2
\end{equation}
and, in terms of partial waves,
\begin{equation}
    V_{\mathrm{ct}}^{(0)}\left({ }^1 S_0\right)=\widetilde{C}_{{ }^1 S_0}=4\pi\left(C_S-3C_T\right),
\end{equation}
\begin{equation}
    V_{\mathrm{ct}}^{(0)}\left({ }^3 S_1\right)=\widetilde{C}_{{ }^3 S_1}=4\pi\left(C_S+C_T\right).
\end{equation}
At NLO, we have
\begin{equation}
    \begin{aligned}
        V_{\mathrm{ct}}^{(2)}\left({\vec p}~', \vec{p}\right)= & C_1 q^2+C_2 k^2+
        \left(C_3 q^2+C_4 k^2\right) \vec{\sigma}_1 \cdot \vec{\sigma}_2 \\
        & +C_5[-i \vec{S} \cdot(\vec{q} \times \vec{k})]+C_6\left(\vec{\sigma}_1 \cdot \vec{q}\right)
        \left(\vec{\sigma}_2 \cdot \vec{q}\right) \\
        & +C_7\left(\vec{\sigma}_1 \cdot \vec{k}\right)\left(\vec{\sigma}_2 \cdot \vec{k}\right),
        \end{aligned}
\end{equation}
and partial-wave decomposition yields
\begin{equation}
    \begin{aligned}
        V_{\mathrm{ct}}^{(2)}\left({ }^1 S_0\right) & =C_{{ }^1 S_0}\left(p^2+p^{\prime 2}\right), \\
        V_{\mathrm{ct}}^{(2)}\left({ }^3 P_0\right) & =C_{{ }^3 P_0} p p^{\prime}, \\
        V_{\mathrm{ct}}^{(2)}\left({ }^1 P_1\right) & =C_{{ }^1 P_1} p p^{\prime}, \\
        V_{\mathrm{ct}}^{(2)}\left({ }^3 P_1\right) & =C_{{ }^3 P_1} p p^{\prime}, \\
        V_{\mathrm{ct}}^{(2)}\left({ }^3 S_1\right) & =C_{{}^3 S_1}\left(p^2+p^{\prime 2}\right), \\
        V_{\mathrm{ct}}^{(2)}\left({ }^3 S_1-{ }^3 D_1\right) & =C_{{ }^3 S_1-{ }^3 D_1} p^2, \\
        V_{\mathrm{ct}}^{(2)}\left({ }^3 D_1-{ }^3 S_1\right) & =C_{{ }^3 S_1-{ }^3 D_1} p^{\prime 2}, \\
        V_{\mathrm{ct}}^{(2)}\left({ }^3 P_2\right) & =C_{{ }^3 P_2} p p^{\prime} .
        \end{aligned}
\end{equation}
The N$^3$LO contacts are
\begin{equation}
    \begin{aligned}
        V_{\mathrm{ct}}^{(4)} & \left({\vec p}~', \vec{p}\right) \\
        = & D_1 q^4+D_2 k^4+D_3 q^2 k^2+D_4(\vec{q} \times \vec{k})^2 \\
        & +\left[D_5 q^4+D_6 k^4+D_7 q^2 k^2+D_8(\vec{q} \times \vec{k})^2\right] \vec{\sigma}_1 \cdot \vec{\sigma}_2 \\
        & +\left(D_9 q^2+D_{10} k^2\right)[-i \vec{S} \cdot(\vec{q} \times \vec{k})] \\
        & +\left(D_{11} q^2+D_{12} k^2\right)\left(\vec{\sigma}_1 \cdot \vec{q}\right)\left(\vec{\sigma}_2 \cdot \vec{q}\right) \\
        & +\left(D_{13} q^2+D_{14} k^2\right)\left(\vec{\sigma}_1 \cdot \vec{k}\right)\left(\vec{\sigma}_2 \cdot \vec{k}\right) \\
        & +D_{15}\left[\vec{\sigma}_1 \cdot(\vec{q} \times \vec{k}) \vec{\sigma}_2 \cdot(\vec{q} \times \vec{k})\right],
        \end{aligned}
\end{equation}
with contributions by partial waves,
\begin{equation}
    \begin{aligned}
        & V_{\mathrm{ct}}^{(4)}\left({ }^1 S_0\right)=\widehat{D}_{{ }^1 S_0}\left(p^{\prime 4}+p^4\right)+D_{{ }^1 S_0} p^{\prime 2} p^2, \\
        & V_{\mathrm{ct}}^{(4)}\left({ }^3 P_0\right)=D_{^3 P_0}\left(p^{\prime 3} p+p^{\prime} p^3\right) \text {, } \\
        & V_{\mathrm{ct}}^{(4)}\left({ }^1 P_1\right)=D_{^1P_1}\left(p^{\prime 3} p+p^{\prime} p^3\right), \\
        & V_{\mathrm{ct}}^{(4)}\left({ }^3 P_1\right)=D_{^3 P_1}\left(p^{\prime 3} p+p^{\prime} p^3\right), \\
        & V_{\mathrm{ct}}^{(4)}\left({ }^3 S_1\right)=\widehat{D}_{^3S_1}\left(p^{\prime 4}+p^4\right)+D_{^3S_1} p^{\prime 2} p^2 \text {, } \\
        & V_{\mathrm{ct}}^{(4)}\left({ }^3 D_1\right)=D_{{ }^3 D_1} p^{\prime 2} p^2, \\
        & V_{\mathrm{ct}}^{(4)}\left({ }^3 S_1-{ }^3 D_1\right)=\widehat{D}_{^3 S_1-{ }^3 D_1} p^4+D_{^3 S_1-{ }^3 D_1} p^{\prime 2} p^2 \text {, } \\
        & V_{\mathrm{ct}}^{(4)}\left({ }^3 D_1-{ }^3 S_1\right)=\widehat{D}_{{ }^3 S_1-{ }^3 D_1} p^{\prime 4}+D_{{ }^3 S_1-{ }^3 D_1} p^{\prime 2} p^2 \text {, } \\
        & V_{\mathrm{ct}}^{(4)}\left({ }^1 D_2\right)=D_{{ }^1 D_2} p^{\prime 2} p^2, \\
        & V_{\mathrm{ct}}^{(4)}\left({ }^3 D_2\right)=D_{{ }^3 D_2} p^{\prime 2} p^2 \text {, } \\
        & V_{\mathrm{ct}}^{(4)}\left({ }^3 P_2\right)=D_{{ }^3 P_2}\left(p^{\prime 3} p+p^{\prime} p^3\right), \\
        & V_{\mathrm{ct}}^{(4)}\left({ }^3 P_2-{ }^3 F_2\right)=D_{^3 P_2-{ }^3 F_2} p^{\prime} p^3, \\
        & V_{\mathrm{ct}}^{(4)}\left({ }^3 F_2-{ }^3 P_2\right)=D_{{ }^3 P_2-{ }^3 F_2} p^{\prime 3} p, \\
        & V_{\mathrm{ct}}^{(4)}\left({ }^3 D_3\right)=D_{{ }^3 D_3} p^{\prime 2} p^2 . \\
        &
        \end{aligned}
\end{equation}
The coefficients $C_{^{(2S+1)}L_J}$ and $C_i$, as well as $D_{^{(2S+1)}L_J}$ and $D_i$, can be transformed into each other through 
invertible matrixs. The specific matrices for these transformations can be found in Ref.~\cite{Chiraleffectivefield}.

\section{Spectral-function representation}

We denote the tensor term of the pion exchange contribution at
NLO (Eq.~\eqref{eq:NLOVT}) as $V^{\mathrm{NLO}}_{T}$. 
Here we demonstrate the process of obtaining its spectral-function representation. 
Our starting point is the potential obtained through dimensional regularization
\begin{equation}
V_T(q)=-\frac{3 g_A^4}{64 \pi^2 f_\pi^4} L(q),
\end{equation}
where the loop function is
\begin{equation}
L(q)=\frac{w}{q} \ln \frac{w+q}{2 m_\pi}.
\end{equation}
The corresponding mass spectrum $\rho_T(\mu)$ (spectral function) is 
\begin{equation}
\rho_T(\mu)=\operatorname{Im} V_T(i \mu)=\frac{3 g_A^4}{128 \pi f_\pi^4} \frac{\sqrt{\mu^2-4 m_\pi^2}}{\mu} .
\end{equation}
Then, the spectral-function representation of this term is obtained through a continuous superposition of Yukawa functions
\begin{equation}
V_T^{SF}=\frac{2}{\pi} \int_{2 M_\pi}^{\tilde{\Lambda}} \mathrm{d} \mu \mu \frac{\rho_T(\mu)}{\mu^2+q^2} \label{eq:SF}
\end{equation}
We do not use the subtracted spectral representation \cite{donoghue1995dispersion} as in Eq.~\eqref{eq:SFR}, 
because the Fourier transform of Eq.~\eqref{eq:SF} directly results in the coordinate space potential
\begin{equation}
\begin{aligned}
\widetilde{V}_T(r)=& -\frac{1}{6 \pi^2 r^3} \int_{2 m_\pi}^{\tilde{\Lambda}} d \mu \mu e^{-\mu r}\left(3+3 \mu r+\mu^2 r^2\right) \\
                   &\times \operatorname{Im} V_T(i \mu).
\end{aligned}
\end{equation}
Thus this part captures the finite range part of the potential in coordinate space
\begin{equation}
V_T^{SF}(q)=-\frac{3 g_A^4}{64 \pi^2 f_\pi^4}\left[ L(\tilde{\Lambda};q)-\ln \frac{1}{2}\left(\frac{\tilde{\Lambda}+s}{m_\pi}\right)\right].
\end{equation}
Then $V^{\mathrm{NLO}}_{T}$ can be divided as
\begin{equation}
    V^{\mathrm{NLO}}_{T}=V^{\mathrm{NLO}}_{T,\mathrm{SF}}+V^{\mathrm{NLO}}_{T,\mathrm{pol}}, \label{eq:NLOVTsep}
\end{equation}
where the remaining polynomial term $V^{\mathrm{NLO}}_{T,\mathrm{pol}}$ is given by $-\frac{3 g_A^4}{64 \pi^2 f_\pi^4}\ln \frac{1}{2}\left(\frac{\tilde{\Lambda}+s}{m_\pi}\right)$.


%
\end{document}